\documentclass[aps,amsmath,amssymb,superscriptaddress,nofootinbib,preprint]{revtex4}
\usepackage{graphicx,bm}

\newcommand{\ba}{\begin{eqnarray}} \newcommand{\ea}{\end{eqnarray}}
\newcommand{\be}{\begin{equation}} \newcommand{\ee}{\end{equation}}
\newcommand{\bw}{\begin{widetext}} \newcommand{\ew}{\end{widetext}}
\newcommand{\nn}{\nonumber}

\newcommand{\arctanh}{\textrm{\,arctanh\,}}
 
\renewcommand{\d}{\textrm{d}} 
\newcommand{\diag}{\textrm{diag}} \newcommand{\e}{\textrm{e}}
 \renewcommand{\i}{\textrm{i}}
 
 \renewcommand{\r}{\rho}
\newcommand{\s}{\sigma} \newcommand{\g}{\gamma}
\newcommand{\eps}{\epsilon} \renewcommand{\l}{\lambda}

 \renewcommand{\bf}{\bfseries}

\newcommand{\fr}[2]{{\textstyle{\frac{#1}{#2}}}}
\newcounter{mycounter}

\begin{document}

\title{Asymmetric inflation: exact solutions}

\author{Roman V. Buniy}
\email{roman@uoregon.edu}
\affiliation{Institute of Theoretical Science, \\ University of Oregon,
Eugene, OR 97403, USA}

\author{Arjun Berera} \email{ab@ph.ed.ac.uk} \affiliation{ School of
Physics, University of Edinburgh, Edinburgh, EH9 3JZ, UK}

\author{Thomas W. Kephart} 
\email{thomas.w.kephart@vanderbilt.edu}
\affiliation{Department of Physics and Astronomy, Vanderbilt
University, Nashville, TN 37235, USA}

\begin{abstract}
We provide exact solutions to the Einstein equations when the Universe
contains vacuum energy plus a uniform arrangements of magnetic fields,
strings, or domain walls.  Such a universe has planar symmetry, i. e.,
it is homogeneous but, not isotropic.  Further exact solutions are
obtained when dust is included and approximate solutions are found for
$w\not=0$ matter. These cosmologies also have planar symmetry. These
results may eventually be used to explain some features in the WMAP
data.  The magnetic field case is the easiest to motivate and has the
highest possibility of yielding reliable constraints on observational
cosmology.
\end{abstract}


\maketitle

\date{today}

\section{Introduction}\label{S:I}

After many successes, standard radiation/matter dominated Big Bang
cosmology was found inadequate to provide solutions to a number of
problems raised when the model was studied in more detail in the light
of modern data. These problems include the horizon problem, the
flatness problem, the magnetic monopole problem, etc. Faced with these
issues, it was clear that a departure from conventional thinking was
required and initial assumptions needed to be
questioned. Specifically, the cosmological constant that was for many
years in disfavor, was reintroduced and gave a solution to Einstein's
equations with exponentially growing scale factor, i.e.,
inflation~{\cite{Guth:1980zm, Linde:1981mu, Albrecht:1982wi,
Linde:1983gd}. This immediately solved the problems listed above,
since it allowed the universe to be in thermal equilibrium, diluted
monopoles, and flattened the curvature. Inflation also allowed quantum
fluctuations in the early universe to expand to super-horizon
sizes. Upon re-entry the fluctuations generate the density
perturbations \cite{Guth:1982ec, Kodama:1985bj, Mukhanov:1990me,
Liddle:1993fq} that led to structure \cite{Peebles, Padmanabhan}.

As more cosmological data became available \cite{Smoot:1992td,
Bennett:1996ce, Kogut:1996us}, more detailed inflationary models have
become necessary to explain it \cite{Lyth:1998xn}. In the last two
decades many inflationary scenarios have been analyzed \cite{Linde,
KT, Dodelson, KR}, but all have one feature in common: homogeneous
isotropic expansion. However, before the onset of inflation, a typical
region of the universe is anything but homogeneous and isotropic. It
is possible that an asymmetric feature in some region could be
stretched out by an asymmetric inflation, but remain imprinted through
the end of inflation, or that a parameter or field initially
asymmetrically distributed, is diluted to a negligible value by the
end of inflation, but with an imprint left on the inflated
universe. Asymmetric features could also be generated by the phase
transition that is responsible for the inflation. We will investigate
these possibilities in models with homogeneous but anisotropic
expansions.

In a previous paper \cite{Berera:2003tf} we gave exact solutions of
Einstein's equations for cases of a universe with planar
symmetry. These include a universe with cosmological constant plus
magnetic fields, cosmic strings or cosmic domain walls aligned
uniformly throughout all space. In this paper we give exact results
that include non-relativistic matter (dust). We also give approximate
solutions for $w\not=0$ matter.

The first year WMAP results \cite{Bennett:2003bz, Spergel:2003cb,
Hinshaw:2003ex} contain interesting large-scale features which warrant
further attention \cite{Tegmark:2003ve},
\cite{deOliveira-Costa:2003pu}. One glaring observational feature is
the suppression of power at large angular scales ($\theta \gtrsim
60^{\circ}$), which is reflected most distinctly in the reduction of
the quadrupole $C_2$. This effect was also seen in the COBE
results~\cite{Smoot:1992td,Kogut:1996us}. After the COBE experiment,
Monte Carlo studies were used to cast doubt on quadrupole
suppression~\cite{Berera:1997wz,Berera:2000wz}, suggesting the effect
could just be statistical. The WMAP analysis~\cite{Bennett:2003bz,
Spergel:2003cb, Hinshaw:2003ex, Tegmark:2003ve,
deOliveira-Costa:2003pu} have arrived at similar
conclusions. Nevertheless, interesting physical effects are not ruled
out, especially since the octupole also appears to be somewhat
suppressed.  Thus it does not seem unreasonable to try to model such
behavior by altering the cosmological model from the standard big bang
plus inflation scenario.  Intriguingly, the more precise measurements
of WMAP also showed that the quadrupole $C_2$ and octupole $C_3$ are
aligned. In particular, the $\ell=2$ and $3$ powers are found to be
concentrated in a plane {\cal P} inclined about $30^{\circ}$ to the
Galactic plane. In a coordinate system in which the equator is in the
plane {\cal P}, the $\ell= 2$ and $3$ powers are primarily in the
$m=\pm \ell$ modes. The axis of this system defines a unique ray and
supports the idea of power in the axial direction being suppressed
relative to the power in the orthogonal plane. These effects seem to
suggest one (longitudinal) direction may have expanded differently
from the other two (transverse) directions, where the transverse
directions describe the equatorial plane {\cal P} mentioned
above. Although this effect once again could be explained away as
statistical~\cite{Bennett:2003bz, Spergel:2003cb, Hinshaw:2003ex,
Tegmark:2003ve, deOliveira-Costa:2003pu}, a realistic physical model
that can explain some or all anisotropic effects in the WMAP data
would be of interest.
 
While there are many ways to approach the issue of global anisotropy
of the Universe, it would be most satisfying to explain global
anisotropy by a simple modification of the conventional
Friedman-Robertson-Walker (FRW) model. To achieve this, one has to
consider an energy-momentum tensor which is spatially non-spherical or
spontaneously becomes non-spherical at each point in space-time. Such
a situation could occur when defects or magnetic fields are
present. Magnetic fields~\cite{Kronberg:1993vk} and cosmic
defects~\cite{Hindmarsh:1994re} can arise in various ways. Moreover,
it is known that large scale magnetic fields exist in the universe,
perhaps up to cosmological
scales~\cite{Kronberg:1993vk,Wick:2000yc}. These considerations
motivate us to focus our attention on the effect magnetic fields and
defects can have on the expansion of the universe.

As a modest step toward understanding the form, significance and
implications of an asymmetric universe, we will modify the standard
spherically symmetric FRW cosmology to a form with only planar
symmetry \cite{Taub:1950ez}. Our choice of the energy-momentum tensor
will result in non-spherical expansion~\footnote{By spherical
expansion we will mean homogeneous isotropic expansion, while in this
paper we will occasionally use the term non-spherical expansion to
describe expansion that has only planar symmetry.} from an initially
spherical symmetric configuration: an initial co-moving sphere will
evolve into a spheroid that can be either prolate or oblate depending
on the choice of matter content.  For the sake of clarity, we first
give some general properties of cosmologies with planar symmetry. (The
universe looks the same from all points but the points all have a
preferred axis.) Our first example will be a universe filled with
dust, uniform magnetic fields and cosmological constant. (Some aspects
of cosmic magnetic fields have been previously studied; see,
e. g. Ref.~\cite{magnetic-fields}.) This is perhaps the most easily
motivated, exactly solvable case to consider and it will give us a
context in which to couch the discussion of other examples with planar
symmetry and cases where planar symmetry is broken. We then describe a
number of other exactly solvable planar symmetric cases.

To set the stage, consider an early epoch in the universe at the onset
of cosmic inflation, where strong magnetic fields have been produced
in a phase transition \cite{Savvidy:1977as, Vachaspati:1991nm,
Enqvist:1994rm, Berera:1998hv}.  Assuming the magnitude of the
magnetic field and vacuum energy ($\Lambda$) densities are initially
about the same, we will find that eventually $\Lambda$ dominates. It
was estimated~\cite{Vachaspati:1991nm} that the initial magnetic field
energy produced in the electroweak phase transition was within an
order of magnitude of the critical density. Other phase transitions
may have even higher initial field values
\cite{Enqvist:1994rm,Berera:1998hv}, or high densities of cosmic
defects. Hence, it is not unphysical to consider a universe with
magnetic fields and $\Lambda$ of comparable magnitudes. If the
magnetic fields are aligned in domains, then some degree of inflation
is sufficient to push all but one domain outside the horizon. (Below
we also discuss the cases where there is one or only a few domains
within the horizon.)

Finally, we should mention that studies of departure from spherical
symmetry and/or departure from standard inflationary cosmology is an
active and controversial area of study. A partial list of topics
includes polarization of light from astrophysical objects and related
phenomena \cite{Birch:1982, Nodland:1997cc, Ralston:2003pf,
Jaffe:2005pw, Hutsemekers:2005iz}, the topology of the universe
\cite{Cornish:1997ab, Luminet:2003dx, Cornish:2003db}, and low $\ell$
mode suppression \cite{Schwarz:2004gk, Kolb:2005me}.  The plan of the
paper follows.

In the second section we review the generalization of a FRW universe
to the case of planar symmetry. We display the Christoffel symbols,
Ricci tensor, and general form of the energy-momentum tensor with the
corresponding Einstein equations and the general form of
energy-momentum conservation.  This section sets up the basic
equations to be solved and also contains a discussion of
thermodynamics in a planar-symmetric universe. We find a natural
splitting of the elements of $T^{\mu\nu}$ into spherically symmetric
and anisotropic pieces. This procedure provides key insight needed to
find exact solutions for the equations of cosmic evolution. In the
third section we carry out a general analysis of the features
resulting from planar symmetry, and give various relations and
inequalities based on energy conditions, many of which involve the
eccentricity of the expansion. Various limiting cases are also
considered.

Sections 4, 5 and 6 treat a universe filled with cosmological
constant, dust and either uniform magnetic fields, aligned cosmic
strings, or aligned cosmic domain walls, respectively. We have given
each of these exactly solvable cases a separate section so that we can
systematically compare and contrast them more easily. Graphics and
limiting cases are both used for this purpose. The magnetic field and
aligned cosmic string cases are qualitatively similar, and both are
substantially different from the case of domain walls. Section 7
contains our conclusions and a brief discussion of how one would apply
our results to density perturbations~\cite{EccIII}. An appendix has
been included to treat the case of matter with generic non-zero choice
for $w$, the parameter that describes the equation of state. These
results are only approximate and so they have been relegated to the
appendix to avoid breaking the flow of the discussion of exact results
presented in the main body of the paper.

\section{Universe with planar symmetry}\label{S:P}

To make the simplest directionally anisotropic universe, we modify the
FRW spherical symmetry of space-time into plane symmetry. (Cylindrical
symmetry is, of course, not appropriate since it introduces preferred
location of the axis of symmetry.) The most general form of a
plane-symmetric metric (up to a conformal transformation)
is~\cite{Taub:1950ez}
\ba(g_{\mu\nu})=\diag{(1,-\e^{2a},-\e^{2a},-\e^{2b})},\label{metric}\ea
where $a$ and $b$ are functions of $t$ and $z$; the $xy$-plane is the
plane of symmetry. We also impose translational symmetry along the
$z$-axis; the functions $a$ and $b$ now depend only on $t$. (Examples
of plane-symmetric spaces include space uniformly filled with either
uniform magnetic fields, static aligned strings, or static stacked
walls, where the defects are at rest with respect to the cosmic
background frame. This situation with defects is artificial or at best
contrived, but could perhaps arise in brane world physics where walls
could be static or walls beyond the horizon could be connected by
static strings. We will not pursue these details here. Of course, any
spherically-symmetric contributions (vacuum energy, matter, radiation)
can be added without altering the planar symmetry.) For the metric
(\ref{metric}), the nonzero Christoffel symbols are \ba
&&\Gamma^0_{11}=\Gamma^0_{22}=\dot{a}\,\e^{2a},\ \ \
\Gamma^0_{33}=\dot{b}\,\e^{2b},\nn\\
&&\Gamma^1_{01}=\Gamma^2_{02}=\dot{a},\ \ \
\Gamma^3_{03}=\dot{b},\nn\ea which results in the following
nonzero-components of the Ricci tensor: \ba
&&{R^0}_0=-(2\ddot{a}+\ddot{b}+2\dot{a}^2+\dot{b}^2),\label{Ricci00}\nn\\
&&{R^1}_1={R^2}_2=-(\ddot{a}+2\dot{a}^2+\dot{a}\dot{b}),\label{Ricci1122}\nn\\
&&{R^3}_3=-(\ddot{b}+\dot{b}^2+2\dot{a}\dot{b}).\label{Ricci33}\nn\ea

To support a symmetry of space-time, the energy-momentum tensor for
the matter has to have the same symmetry. In the case of planar
symmetry this requires \ba ({T^\mu}_\nu)=(8\pi G)^{-1}
\diag(\xi,\eta,\eta,\zeta).\label{T}\ea Here the energy density $\xi$,
transverse $\eta$ and longitudinal $\zeta$ tension densities are
functions only of time.  The corresponding Einstein equations are
\ba&&\dot{a}^2+2\dot{a}\dot{b}=\xi,\label{A1}\\
&&\ddot{a}+\ddot{b}+\dot{a}^2+\dot{a}\dot{b}+\dot{b}^2=\eta,\label{A2}\\
&&2\ddot{a}+3\dot{a}^2=\zeta.\label{A3}\ea We also need the equation
expressing covariant conservation of the energy-momentum [a direct
consequence of Eqs.~(\ref{A1})--(\ref{A3})]: \ba
\dot{\xi}+2\dot{a}(\xi-\eta)+\dot{b}(\xi-\zeta)=0.
\label{A4}\ea

\begin{table}[htb]
\caption{\label{table}The components of the energy momentum (\ref{T})
for various contributions to the matter. In the text, tables and
figure captions we use the labels $\Lambda$, $w$, M, S, and W to
represent cosmological constant, matter with equation of state
$\rho=wp$, magnetic fields, strings, and walls,
respectively. Occasionally we label dust with the symbol $0$.  }
\begin{ruledtabular}
\begin{tabular}{lp{1cm}cp{0.5cm}cp{0.5cm}c}
& & $\xi$ & & $\eta$ & & $\zeta$\\ \hline vacuum energy ($\Lambda$) &
& $\l$ & & $\l$ & & $\l$\\ matter ($w$) & & $\rho$ & & $-w\rho$ & &
$-w\rho$\\ magnetic field (M) & & $\eps$ & & $-\eps$ & & $\eps$\\
strings (S) & & $\eps$ & & $0$ & & $\eps$\\ walls (W) & & $\eps$ & &
$\eps$ & & $0$
\end{tabular}
\end{ruledtabular}
\end{table}

We consider now the thermodynamics of cosmological models evolving
anisotropically. Energy density $\rho$ and pressure $p$ correspond to
the spherically-symmetric part of the energy-momentum tensor,
$\xi=\lambda+\rho+\tilde{\xi}$, $\eta=\lambda-p+\tilde{\eta}$,
$\zeta=\lambda-p+\tilde{\zeta}$, where we have written tildes on the
anisotropic parts of the energy-momentum tensor. As in the isotropic
case~\cite{Weinberg}, we have \ba T\d p/\d T=\rho+p,\ea where $T$ is
the temperature. Up to an additive constant, the entropy in a volume
$V$ is \ba S=(\rho+p)V/T.\ea Taking $V=V_\i\,\mathrm{e}^{2a+b}$ we
find \ba\dot{S}/S=2\dot{a}+\dot{b}+\dot{\rho}/(\rho+p).\label{dotS}\ea
For an adiabatic process, the entropy in a comoving volume is
conserved; thus the right-hand side of Eq.~(\ref{dotS}) vanishes, \ba
\dot{\rho}+(2\dot{a}+\dot{b})(\rho+p)=0.\label{A5}\ea For matter with
the equation of state $p=w\rho$, Eq.~(\ref{A5}) gives \ba
\rho=\rho_\i\,\e^{-(1+w)(2a+b)}.\label{A5s}\ea Eq.~(\ref{A5})
expresses covariant conservation of the isotropic part of the
energy-momentum. Since the total energy-momentum is conserved locally
[Eq.~(\ref{A4})], the same holds for its anisotropic part, \ba
\dot{\tilde{\xi}} +2\dot{a}(\tilde{\xi}-\tilde{\eta})
+\dot{b}(\tilde{\xi}-\tilde{\zeta})=0.\label{A6}\ea Eq.~(\ref{A6})
will be the key to finding our exact solutions.

\section{General properties}\label{S:G}

Before considering specific models for the energy-momentum, we first
establish several general features of an anisotropic universe
described by Eqs.~(\ref{A1})--(\ref{A4}). These results will be
important for conceptual understanding of solutions and asymptotics
bounding the effects caused by asymmetry and comparing the results for
various types of asymmetric components.

\begin{list}{\bf \arabic{mycounter}.}{\usecounter{mycounter} 
\setlength{\parsep}{0ex}\setlength{\leftmargin}{0cm}
\setlength{\itemindent}{\leftmargin}\addtolength{\itemindent}{\labelwidth}
\addtolength{\itemindent}{\labelsep} }

\item\label{i:shape} We assume that before anisotropic effects became
important, the universe had expanded isotropically. During the initial
phase of anisotropic expansion, when anisotropic contributions to the
energy-momentum are significant, different tension densities in the
longitudinal and transverse directions cause comoving spheres to
evolve into spheroids. At a later phase, when all contributions except
for the vacuum energy fade away, longitudinal and transverse expansion
rates become equal and the expansion proceeds isotropically. Thus, in
this universe, each initial sphere develops an eccentricity; whether
the resulting spheroid is oblate or prolate depends on which tension
dominated during the initial phase of deformation.

\item\label{i:dot-a} We assume that initially~\footnote{The subscript
``$\i$'' refers to the moment of transition from isotropic to
anisotropic dynamics.} space is isotropic, $a_\i=b_\i$, and is
expanding isotropically, $\dot{a}_\i=\dot{b}_\i>0$.  Without loss of
generality we set $a_\i=0$, which is equivalent to a simple rescaling
of scale factors $\e^a$ and $\e^b$. For expansion in some direction to
change into contraction we need to cross the point where the expansion
rate in this direction becomes zero. Since the energy density is
positive, from Eq.~(\ref{A1}) it follows that there can be no
contraction in the transverse direction, $\dot{a}>0$. (See
Figs.~\ref{fig-m-a}, \ref{fig-s-a}, \ref{fig-w-a}.)

\item\label{i:eta-zeta} If the transverse tension is always smaller
(larger) than the longitudinal tension, then initially spherical
region of space-time has expanded asymptotically to the shape of an
oblate (prolate) spheroid. Indeed, if $\eta<\zeta$, then
Eqs.~(\ref{A2}) and (\ref{A3}) give
$\ddot{a}-\ddot{b}<-2\dot{a}^2+\dot{a}\dot{b}+\dot{b}^2$. Consider two
cases: (i) when $\dot{a}\le\dot{b}$, we arrive at
$\ddot{a}-\ddot{b}>0$, which upon integration leads to a
contradiction, $\dot{a}-\dot{b}>0$; (ii) when $\dot{a}>\dot{b}$, we
find $\dot{a}-\dot{b}>2\int_{t_\i}^t\d t(\dot{b}^2-\dot{a}^2)$, which
is allowed. Similarly, if $\eta>\zeta$, then $\dot{a}\ge\dot{b}$ leads
to a contradiction, while $\dot{a}<\dot{b}$ is allowed. In the allowed
cases, further integration leads to the result: $a>b$ when
$\eta<\zeta$, and $a<b$ when $\eta>\zeta$. (See
Figs.~\ref{fig-m-eccentricity-new}, \ref{fig-m-ab-new},
\ref{fig-s-eccentricity-1}, \ref{fig-s-ab},
\ref{fig-w-eccentricity-1}, \ref{fig-w-ab} for examples of different
cases.)

\item\label{i:dot-b} On the other hand, longitudinal contraction is
possible. To find a moment $t=t_*$ when longitudinal expansion changes
into contraction, we set $\dot{b}_*=0$ in Eqs.~(\ref{A1})--(\ref{A3})
and find $\ddot{b}_*=\eta_*+\fr{1}{2}(\xi_*-\zeta_*)$. Examining
entries in Table~\ref{table} we see that, except for the case when the
magnetic field dominates~\footnote{When we say that some contribution
to the energy-momentum dominates we mean by this that the contribution
is the largest for \emph{all} times.}, $\ddot{b}_*$ is positive. It
follows that only in the magnetic-field-dominating case, can space
contract in the longitudinal direction; see Figs.~\ref{fig-m-b},
\ref{fig-s-b}, \ref{fig-w-b} for examples of the different cases. If
the initial magnetic field is sufficiently strong, the longitudinal
size can become smaller then its initial value (curves with $b<0$ in
Fig.~\ref{fig-m-b}). However, no matter how strong the initial field
is, after a sufficiently long period of time contraction turns into
expansion following the correct asymptotic behavior.

\item\label{i:dot-xi} For all known forms of matter the components of
the energy-momentum tensor satisfy the dominant energy
condition~\cite{energy-momentum,nec}; in our case it says $\xi\ge 0$,
$\xi\ge\eta$, and $\xi\ge\zeta$ (see Table~\ref{table} for
examples). Evaluating Eq.~(\ref{A4}) at the initial time, we find
$\dot{\xi}_\i=-\dot{a}_\i(3\xi_\i-2\eta_\i-\zeta_\i)$; from the energy
conditions it now follows that the energy density does not increase
initially, $\dot{\xi}_\i\le 0$. Let us investigate whether the energy
can increase at a later time. For this to happen, we need to have
$\dot{\xi}_*=0$ for some $t=t_*$; Eqs.~(\ref{A1}) and (\ref{A4}) then
give $\dot{a}_*^2(4\eta_*-3\xi_*-\zeta_*) =\xi_*(\xi_*-\zeta_*)$. This
equation has a real solution for $\dot{a}_*$ only when the wall
contribution dominates; see Table~\ref{table}. However, even in the
worst case, with only the wall contribution, the energy density cannot
increase. Indeed, using $\xi=\eta$, we find $\dot{a}_*^2=\xi_*$. Since
$\dot{a}>0$ by {\bf \ref{i:dot-a}}, Eq.~(\ref{A1}) gives
$\dot{b}_*=0$. In {\bf \ref{i:dot-b}} we found that, unless magnetic
field contribution dominates, $\ddot{b}_*>0$ and thus for walls
$\dot{b}$ cannot become zero if initially $\dot{b}_\i>0$. This proves
that $\dot{\xi}\le 0$ in all cases.

\item\label{i:dot-a-max} Let us prove that the transverse expansion
rate has its maximum at $t=t_\i$. First, from Eqs.~(\ref{A1}) and
(\ref{A3}) it follows that $\ddot{a}_\i=\fr{1}{2}(\zeta_\i-\xi_\i)$
and so $\ddot{a}_\i\le 0$ because of the energy conditions in {\bf
\ref{i:dot-xi}}. It follows that $\dot{a}$ initially decreases with
time. Suppose, however, that $\dot{a}$ starts increasing and reaches
its initial value $\dot{a}_\i$ for the first time at the moment
$t_*$. Equation~(\ref{A3}) then gives
$\ddot{a}_*=\fr{1}{2}(\zeta_*-\xi_\i)$ and $\xi_\i\ge\xi_*\ge\zeta_*$
from {\bf \ref{i:dot-xi}} leads to $\ddot{a}_*\le 0$. This is
impossible since to reach the first point at which
$\dot{a}_*=\dot{a}_\i$ we need to have $\ddot{a}_*>0$. We thus
conclude $\dot{a}\le\dot{a}_\i$, which was to be demonstrated. (See
Figs.~\ref{fig-m-a}, \ref{fig-s-a}, \ref{fig-w-a}.)

\item\label{i:dot-b-max} Consider now longitudinal expansion. From
Eqs.~(\ref{A1})--(\ref{A3}) we have
$\ddot{b}_\i=\eta_\i-\fr{1}{2}(\xi_\i+\zeta_\i)$. Examining entries in
Table~\ref{table} we see that $\ddot{b}_\i\le 0$ except for the case
when the wall contribution dominates. Excluding this case, we repeat
the argument in {\bf \ref{i:dot-a-max}} and for the point at which
$\dot{b}_*=\dot{b}_\i$ we find $\ddot{b}_*
=\eta_*-\fr{1}{2}(\xi_*+\zeta_*)+\dot{a}_*^2-\dot{a}_\i^2$. Using the
result in {\bf \ref{i:dot-a-max}}, we have $\ddot{b}_*
\le\eta_*-\fr{1}{2}(\xi_*+\zeta_*)\le 0$, which contradicts
$\ddot{b}_*>0$ that we need to make $\dot{b}_*=\dot{b}_\i$. We thus
conclude that, except when walls contribution dominates, the
longitudinal expansion rate has its maximum at $t=t_\i$. (See
Figs.~\ref{fig-m-b}, \ref{fig-s-b}, \ref{fig-w-b}.)

\item\label{i:2ab-max} Next we derive a bound on the spacial
volume. Consider two cases: (i) when $\dot{a}\le\dot{b}$,
Eq.~(\ref{A1}), $\dot{\xi}\le 0$ in {\bf \ref{i:dot-xi}}, and
$\dot{a}>0$ in {\bf \ref{i:dot-a}} lead to
$\ddot{a}(1+\dot{b}/\dot{a})+\ddot{b}\le 0$ and so
$2\ddot{a}+\ddot{b}\le 0$; (ii) when $\dot{a}\ge\dot{b}$,
Eq.~(\ref{A1}), $\dot{\xi}\le 0$ in {\bf \ref{i:dot-xi}}, and
$\dot{b}>0$ in {\bf \ref{i:dot-b}} (which is true unless magnetic
field contribution dominates) lead to
$\ddot{a}(1+\dot{a}/\dot{b})+\ddot{b}\dot{a}/\dot{b}\le 0$ and so
$2\ddot{a}+\ddot{b}\le 0$ again. In both cases integration gives the
following bound for the volume of space: \ba
2a+b\le(3\xi_{\i})^{\frac{1}{2}}(t-t_\i).\ea

\item\label{i:ab-max} The energy condition $\xi\ge\zeta$ bounds the
eccentricity from above. Indeed, from Eqs.~(\ref{A1}) and (\ref{A3})
follows $\dot{a}-\dot{b} =(\zeta-\xi-2\ddot{a})/2\dot{a}$. Integrating
and using $\dot{a}>0$ from {\bf \ref{i:dot-a}} and $\xi\ge\zeta$ from
{\bf \ref{i:dot-xi}} we arrive at the following bound: \ba
\e^{a-b}\le\dot{a}_{\i}/\dot{a}.\label{ab-ineq1}\ea From {\bf
\ref{i:dot-a-max}} we have $\dot{a}_{\i}/\dot{a}\ge 1$ and so
Eq.~(\ref{ab-ineq1}) allows $a>b$. Solving Eq.~(\ref{A3})
asymptotically for large $t$, we find
$\dot{a}\sim(\zeta/3)^{\frac{1}{2}}$, which turns Eq.~(\ref{ab-ineq1})
into an asymptotic bound
\begin{eqnarray}
\e^{a-b}\lesssim(\xi_\i/\zeta)^{\frac{1}{2}}.\label{ab-ineq1a}
\end{eqnarray}
Since $\e^{a-b}$ is a convenient variable, we define it to be the
``pseudo-eccentricity''~\footnote{The standard definition of the
eccentricity of an ellipse, with semi-major axis of length $A=e^a$,
and semi-minor axis of length $B=e^b$, is
$\frac{\sqrt{A^2-B^2}}{B}=\sqrt{e^{2(a-b)}-1}$. We are interested in
spheroids that can be either prolate or oblate. If a cross section
that is tangent to the symmetry axis of the spheroid is an ellipse
with axes $A$ along the symmetry axis and $B$ normal to that axis,
then either one can be larger. This distinction is not contained in
the definition of eccentricity, so a more appropriate measure for our
purposes is the ratio $\frac{A}{B}=e^{a-b}$ which we will call the
pseudo-eccentricity.}.

\item\label{i:xi-zeta} When $\xi=\zeta$ (the vacuum energy with any
combination of magnetic fields and strings aligned in the same
direction), similarly to {\bf \ref{i:ab-max}} we find \be
\e^{a-b}\sim(\xi_\i/\zeta) ^{\frac{1}{2}}.\label{A:ab-asymptotic}\ee

\item\label{i:ab-min} We consider here the case when neither the
magnetic field nor the wall contribution dominates. Using $\dot{b}>0$
from {\bf\ref{i:dot-b}} and $\dot{b}\le\dot{a}_\i$ from
{\bf\ref{i:dot-b-max}}, Eqs.~(\ref{A1}) and (\ref{A2}) together with
the condition $\xi\ge\eta$ from {\bf\ref{i:dot-xi}} lead to
$\dot{a}-\dot{b}\ge(\ddot{a}+\ddot{b})/\dot{b}$, and integration gives
the bound \ba \e^{a-b}\ge(\dot{b}/\dot{a}_{\i})
\,\e^{\dot{a}/\dot{a}_{\i}-1}.\label{ab-ineq2}\ea The right-hand side
of Eq.~(\ref{ab-ineq2}) does not exceed unity, so $a<b$ is
allowed. Using now asymptotic expressions for $\dot{a}$ and $\dot{b}$
(which are obtained from Eqs.~(\ref{A1}) and (\ref{A3})), we find \ba
\e^{a-b}\gtrsim\left[\frac{3\xi-\zeta}{2(\xi_{\i}\zeta)
^{\frac{1}{2}}}\right]\,
\exp{\left[(\zeta/\xi_{\i})^{\frac{1}{2}}-1\right]}.\label{ab-ineq2a}\ea
Asymptotically $\xi,\zeta\sim\lambda$, and so the bounds
(\ref{ab-ineq1a}) and (\ref{ab-ineq2a}) bracket the
pseudo-eccentricity as follows: \ba(\l/\xi_{\i})^{\frac{1}{2}}\,
\exp{\left[(\l/\xi_{\i})^{\frac{1}{2}}-1\right]}\lesssim
\e^{a-b}\lesssim (\xi_{\i}/\l)^{\frac{1}{2}}\label{ab-ineq12a}.\ea The
two bounds in Eq.~(\ref{ab-ineq12a}) do not contradict each other
since $\xi_{\i}\ge\l$.

\item\label{i:xi-eta} When $\xi=\eta$ (the vacuum energy with walls)
we have $\dot{a}-\dot{b}=(\ddot{a}+\ddot{b})/\dot{b}$. Using now
$\ddot{a}\le-\fr{1}{2}\ddot{b}$ from {\bf\ref{i:2ab-max}} and
integrating we find
$\e^{a-b}\le(\dot{b}/\dot{a}_\i)^{\frac{1}{2}}$. Asymptotically,
comoving spheres evolve into prolate ellipsoids, \ba
\e^{a-b}\lesssim(\l/\xi_\i)^{\frac{1}{4}}.\ea

\item\label{i:a2b-min} By {\bf \ref{i:dot-a-max}}, the transverse
expansion rate has its maximum at $t=t_{\i}$. Also, for all physical
matter contributions, the energy density exceeds the vacuum energy
(see Table~\ref{table}). From Eq.~(\ref{A1}) we then have
$\dot{a}+2\dot{b}\ge \l/\dot{a}$, and thus
\begin{eqnarray}
a+2b\ge\l(3/\xi_{\i})^{\frac{1}{2}}(t-t_{\i}).\label{a2b-ineq}
\end{eqnarray}

\item\label{i:a-min} Except when contribution of matter with $w>0$
dominates, the longitudinal tension exceeds the vacuum
energy. Equation~(\ref{A3}) together with the condition
$\dot{a}\le\dot{a}_{\i}$ from {\bf \ref{i:dot-a-max}} then gives \ba
a\ge\l(3\xi_{\i})^{-\frac{1}{2}}(t-t_{\i}).\ea

\item\label{i:limits} When $\lambda>0$, we have asymptotics $\xi$,
$\eta$, $\zeta\sim\l$ and $\dot{a}$,
$\dot{b}\sim(\fr{1}{3}\l)^{\frac{1}{2}}$. In order to find asymptotics
when $\lambda=0$, we first observe that all quantities have power law
behavior: $\xi\sim C_\xi t^{-s_\xi}$, etc. Equation~(\ref{A3}) gives
$s_\zeta=2$, $s_{\dot{a}}=1$. From Eqs.~(\ref{A2}) and (\ref{A3}) we
then have either $s_{\dot{b}}\ge 1$, $s_\xi=s_\eta=2$ or
$s_{\dot{b}}<1$, $s_\xi=s_{\dot{b}}+1$,
$s_\xi=2s_{\dot{b}}$. Examining entries in Table~\ref{table}, we
conclude that from $s_\zeta=2$ it follows that
$\min(s_\eps,s_\rho)=2$; if there is more than one anisotropic
component, then $s_\eps$ is the smallest anisotropic exponent. This
results in $s_\xi=s_\eta=2$, $s_{\dot{b}}\ge 1$. (See
Table~\ref{table:summary} for examples.)

\item\label{i:limits-approach} Let us find how asymptotics for
$\lambda>0$ in {\bf \ref{i:limits}} approach their limiting
values. From Eq.~(\ref{A4}) we have \ba{(\delta\xi)}\,\dot{}
+\left(\fr{1}{3}\l\right)^{\frac{1}{2}}(3-2n_\eta-n_\zeta)
\delta\xi=0,\label{d-xi}\ea where $n_\eta=\delta\eta/\delta\xi$ and
$n_\zeta=\delta\zeta/\delta\xi$. When $n_\eta$ and $n_\zeta$ are
constants (for example, when, besides the vacuum energy, there is only
one other contribution from Table~\ref{table}) or slowly varying
functions, Eq.~(\ref{d-xi}) gives $\delta\xi\propto \e^{-t/t_\xi}$
with characteristic time \ba
t_\xi=(\fr{1}{3}\l)^{-\frac{1}{2}}(3-2n_\eta-n_\zeta)^{-1}.\ea For the
cases $\Lambda\text{M}$, $\Lambda\text{S}$ and $\Lambda\text{W}$,
$(\fr{1}{3}\l)^{\frac{1}{2}}t_\xi$ equals $\fr{1}{4}$, $\fr{1}{2}$,
and $1$, respectively. Similarly, from Eqs.~(\ref{A1})--(\ref{A3}) we
obtain
\begin{eqnarray}
&&{(\delta\dot{a})}\,\dot{}+\left(\fr{1}{3}\l\right)^{\frac{1}{2}}
\left[(3-4n_\zeta)\delta\dot{a}-2n_\zeta\delta\dot{b}\right]=0,
\label{d-a-dot}\\
&&{(\delta\dot{b})}\,\dot{}+\left(\fr{1}{3}\l\right)^{\frac{1}{2}}
\left[(4n_\zeta-4n_\eta)\delta\dot{a}+(3-2n_\eta+2n_\zeta)
\delta\dot{b}\right]=0,\label{d-b-dot}
\end{eqnarray} 
Solving these equations and keeping only the leading terms, we find
$\delta\dot{a}\propto t \e^{-t/t_{\dot{a}}}$, $\delta\dot{b}\propto
t\e^{-t/t_{\dot{b}}}$ for the case $\Lambda\text{M}$ with
$(\fr{1}{3}\l)^{\frac{1}{2}}t_{\dot{a}}
=(\fr{1}{3}\l)^{\frac{1}{2}}t_{\dot{b}}=\fr{1}{3}$, and
$\delta\dot{a}\propto t \e^{-t/t_{\dot{a}}}$, $\delta\dot{b}\propto
t\e^{-t/t_{\dot{b}}}$ for the cases $\Lambda\text{S}$ and
$\Lambda\text{W}$ with $(\fr{1}{3}\l)^{\frac{1}{2}}t_{\dot{a}}$ equal
to $1$ and $\fr{1}{3}$, respectively, with
$(\fr{1}{3}\l)^{\frac{1}{2}}t_{\dot{b}}$ equal to $1$ in both
cases. (See Figs.~\ref{fig-m-a}, \ref{fig-m-b}, \ref{fig-s-a},
\ref{fig-s-b}, \ref{fig-w-a}, \ref{fig-w-b}.)

\end{list}

For the reader's convenience we have collected in Table~\ref{summary}
many of the general results proved in
{\bf\ref{i:shape}}--{\bf\ref{i:limits-approach}}, conditions of their
applicability, and examples of matter content when the results can be
used.

\begin{table}[htb]
\caption{\label{summary}Summary of general results concerning the
system described by Eqs.~(\ref{A1})--(\ref{A4}). Initially, space is
assumed to be isotropic, $a_\i=b_\i=0$, and expanding isotropically,
$\dot{a}_\i=\dot{a}_\i>0$. In derivations of some of the results the
energy conditions, $\xi\ge 0$, $\xi\ge\eta$, $\xi\ge\zeta$, were
assumed. The symbols in the last column indicate the matter content
types for which the corresponding result in the second column is
applicable. If there are more than two components, then their
arbitrary combination is allowed. Also, the result can be extended to
matter content which is not indicated in the forth column provided
that it does not exceed the indicated components.}
\begin{ruledtabular}
\begin{tabular}{cp{1cm}cp{1cm}cp{1cm}c}
No.& & Result & & Conditions & & Applicable to \\ \hline
{\bf\ref{i:dot-a}} & & $\dot{a}>0$ & & & & $\Lambda$, M, S, W, $w$
\\{\bf\ref{i:eta-zeta}a} & & $a>b$ & & $\eta<\zeta$ & & M, S \\
{\bf\ref{i:eta-zeta}b} & & $a<b$ & & $\eta>\zeta$ & & W \\
{\bf\ref{i:dot-b}} & & $\dot{b}>0$ & & $\xi+2\eta-\zeta\ge 0$ & &
$\Lambda$, S, W, $w$ \\ {\bf\ref{i:dot-xi}} & & $\dot{\xi}\le 0$ & & &
& $\Lambda$, M, S, W, $w$\\ {\bf\ref{i:dot-a-max}} & &
$\dot{a}\le\dot{a}_\i$ & & & & $\Lambda$, M, S, W, $w$\\
{\bf\ref{i:dot-b-max}} & & $\dot{b}\le\dot{a}_\i$ & &
$\xi-2\eta+\zeta\ge 0$ & & M, S, $w$ \\ {\bf\ref{i:2ab-max}} & &
$\max{(2a+b)}$ & & $\xi+2\eta-\zeta\ge 0$ & & $\Lambda$, S, W, $w$ \\
{\bf\ref{i:ab-max}} & & $\max{(a-b)}$ & & & & $\Lambda$, M, S, W,
$w$\\ {\bf\ref{i:xi-zeta}} & & $a-b$ & & $\xi=\zeta$ & & $\Lambda$, M,
S \\ {\bf\ref{i:ab-min}} & & $\min{(a-b)}$ & & $\xi-2\eta+\zeta\ge 0$,
$\xi+2\eta-\zeta\ge 0$ & & S, $w$ \\ {\bf\ref{i:xi-eta}} & &
$\max{(a-b)}$ & & $\xi=\eta$ & & $\Lambda$, W \\ {\bf\ref{i:a2b-min}}
& & $\min{(a+2b)}$ & & & & $\Lambda$, M, S, W, $w$ \\
{\bf\ref{i:a-min}} & & $\min{a}$ & & $\zeta\ge\l$ & & $\Lambda$, M, S,
W
\end{tabular}
\end{ruledtabular}
\end{table}

\section{Magnetic fields}\label{S:B}

We are now in a position to extend the analysis of a universe with
cosmological constant plus magnetic fields of
Ref.~\cite{Berera:2003tf} to the more general case that also includes
dust. The analysis is similar to the case without dust, but adds to it
a new variable and requires the use of Eq.~(\ref{A5}). We will find
exact solutions in this case, and in the cases where magnetic field is
replaced by strings or walls. Lest the reader become complacent with
the ease at which these solutions have been found, we will show that
if we have instead $\Lambda+\text{M/S/W}+w$, the resulting
differential equations are much more complicated and difficult to
solve. Numerical techniques are still available by which we will solve
these equations and present the results graphically.

In the case of cosmological constant, magnetic fields and $w\not=0$
matter, Eqs.~(\ref{A1}), (\ref{A3}) and (\ref{A6}) with the
corresponding entries from Table~\ref{table} give
\begin{eqnarray}
&&\dot{a}^2+2\dot{a}\dot{b}=\l+\rho+\eps,\label{B:1}\\
&&2\ddot{a}+3\dot{a}^2=\l-w\rho+\eps,\label{B:3}\\
&&\dot{\eps}+4\dot{a}\eps=0.\label{B:6}
\end{eqnarray}
From the conservation of the anisotropic part of the energy-momentum,
Eq.~(\ref{B:6}), we find
\begin{eqnarray}
a=\fr{1}{4}\ln{(\eps_\i/\eps)}.\label{B:a0}
\end{eqnarray}
Substituting this result into Eq.~(\ref{B:3}), we arrive at
\begin{equation}
\eps\ddot{\eps}-\fr{11}{8}\dot{\eps}^2+2\eps^2(\l-w\rho+\eps)=0.\label{B:e-eq}
\end{equation}
Eq.~(\ref{B:e-eq}) does not explicitly involve the independent
variable $t$. To use this fact, we let $\eps$ be the independent
variable and introduce a new dependent variable
$f=\fr{1}{2}\dot{\eps}^2$. After this change Eq.~(\ref{B:e-eq}) becomes
\begin{equation}
\eps f'-\fr{11}{4}f+2\eps^2(\l-w\rho+\eps)=0;\label{B:f-eq}
\end{equation}
here the prime means differentiation with respect to $\eps$. Solving
Eq.~(\ref{B:1}) for $\dot{b}$ and substituting it into Eq.~(\ref{A5})
we find (after some algebra)
\begin{equation}
\eps f\rho'-(1+w)\rho\left[\fr{3}{8}f
+\eps^2(\l+\rho+\eps)\right]=0.\label{B:rho-eq}
\end{equation}
The system of coupled differential equations~(\ref{B:f-eq}) and
(\ref{B:rho-eq}) can be replaced by the following system of coupled
integral equations:
\begin{eqnarray}
&&f=\fr{8}{3}\l\eps^2+\fr{8}{3}\eps_\i^{-\frac{3}{4}}(\rho_\i+4\eps_\i)
\eps^{\frac{11}{4}}-8\eps^3-2w\eps^{\frac{11}{4}}
\int_\eps^{\eps_\i}\d\eps\,\rho\eps^{-\frac{7}{4}}.\label{B:f}\\
&&\rho=\rho_\i(\eps/\eps_\i)^{\frac{3}{8}(1+w)}\psi
\left[1+(1+w)\rho_\i\eps_\i\int_\eps^{\eps_\i}\d\eps\,
(\eps/\eps_\i)^{1+\frac{3}{8}(1+w)}\psi
f^{-1}\right]^{-1},\label{B:rho}
\end{eqnarray}
where
\begin{equation} 
\psi=\exp{\left[-(1+w)\int_\eps^{\eps_\i}\d\eps\,\eps(\l+\eps)f^{-1}\right]}. 
\label{B:psi}
\end{equation}

We were not able to find the exact solution to the coupled system of
Eqs.~(\ref{B:f-eq}) and (\ref{B:rho-eq}) for arbitrary $w$. Instead,
in the remainder of this section we derive the exact solution for
$w=0$ and defer derivation of an approximate solution for arbitrary
$w$ until Appendix~\ref{S:B0}. (Another exactly solvable case $w=-1$
is not really a separate case since it can be obtained from the
solution for $w=0$ by setting $\rho=0$ and redefining $\l$.)

For $w=0$, Eqs.~(\ref{B:f-eq}) and (\ref{B:rho-eq}) give
\begin{eqnarray}
&&f=\fr{8}{3}\l\eps^2+\fr{8}{3}\eps_\i^{-\frac{3}{4}}(\rho_\i+4\eps_\i)
\eps^{\frac{11}{4}}-8\eps^3,\label{B:f0}\\
&&\rho=\rho_\i(\eps/\eps_\i)^\frac{3}{4}\left[1+F(\eps/\eps_\i)\right]^{-1}
\left[\frac{\l
+(\rho_\i+4\eps_\i)(\eps/\eps_\i)^\frac{3}{4}-3\eps}{\l+\rho_\i+\eps_\i}
\right]^{-\frac{1}{2}},
\label{B:rho0}
\end{eqnarray}
where
\begin{eqnarray}
F(\eps/\eps_\i)
=\fr{3}{8}(\rho_\i/\eps_\i)[1+(\l+\rho_\i)/\eps_\i]^{\frac{1}{2}}
\int_{\eps/\eps_\i}^{1}\d x\,x^{-\frac{1}{4}}
\left\{\l/\eps_\i+(4+\rho_\i/\eps_\i)x^{\frac{3}{4}}-3x\right\}^{-\frac{3}{2}}.
\label{B:F}
\end{eqnarray}
From Eqs.~(\ref{A5s}), (\ref{B:a0}) and (\ref{B:rho0}) we find
\begin{eqnarray}
b=\fr{1}{2}\ln\frac{\l+(\rho_\i+4\eps_\i)(\eps/\eps_\i)^\frac{3}{4}-3\eps}
{\l+\rho_\i+\eps_\i}
-\fr{1}{4}\ln{(\eps/\eps_\i)}+\ln\left[1+F(\eps/\eps_\i)\right].
\label{B:b0}
\end{eqnarray}
Finally, time dependence of the above functions $\rho(\eps)$,
$a(\eps)$ and $b(\eps)$ can be deduced from the function $\eps(t)$,
which is given implicitly by
\begin{eqnarray}
t-t_\i=\fr{1}{4}
\int^{\eps_\i}_{\eps}\d\eps\left[
\fr{1}{3}\l\eps^2+\fr{1}{3}\eps_\i^{-\frac{3}{4}} (4\eps_\i+\rho_\i)
\eps^{\frac{11}{4}}-\eps^3\right]^{-\frac{1}{2}}\label{B:integral0}
\end{eqnarray}
as it follows from $f=\fr{1}{2}\dot{\eps}^2$ and Eq.~(\ref{B:f0}).

From a physical point of view, the value of $\e^{a-b}$ is the largest
for the most anisotropic case; for fixed $\eps_\i$, this is achieved
for $\rho_\i=0$ when $\e^{a-b}\sim(1+\eps_\i/\l)^{\frac{1}{2}}\ge
1$. The case of infinitely large $\rho_\i$ corresponds to the most
isotropic case when $\e^{a-b}=1$. It follows that $\e^{a-b}\ge 1$ for
any $\rho_\i$. A careful inspection of the solution embodied in
Eqs.~(\ref{B:a0}), (\ref{B:b0}) and (\ref{B:integral0}) confirms this
and shows that the space is oblate and its pseudo-eccentricity
monotonically increases from its initial value (unity) to its
asymptotic value. More magnetic field increases the anisotropy; more
matter reduces anisotropy, but neither can change an oblate ellipsoid
into a prolate one.

In the case $w=0$, asymptotics from Appendix~\ref{S:B0} simplify as
follows:
\begin{eqnarray}
&&\eps\sim
\eps_\i\,\exp\left[-4(\l/3)^{\frac{1}{2}}(t-t_\i+\tau)\right],
\label{B:e-asympt}\\ &&a\sim(\l/3)^{\frac{1}{2}}(t-t_\i+\tau), 
\label{B:a-asympt}\\ &&b\sim (\l/3)^{\frac{1}{2}}(t-t_\i+\tau)-\fr{1}{2}
\ln{\left[1+(\rho_\i+\eps_\i)/\l\right]}+\ln{[1+F(0)]},
\label{B:b-asympt}\\ && \rho\sim\rho_\i 
\left[1+(\l+\rho_\i)/\eps_\i\right]^{\frac{1}{2}}[1+F(0)]^{-1}
\exp\left[-(3\l)^{\frac{1}{2}}(t-t_\i+\tau)\right],\label{B:rho-asympt}
\end{eqnarray}
where
\begin{eqnarray}
\tau=\fr{1}{4}\int_{0}^{\eps_\i}
\d\eps\left\{\left(\fr{1}{3}\l\eps^2\right)^{-\frac{1}{2}}
-\left[\fr{1}{3}\l\eps^2+\fr{1}{3}\eps_\i^{-\frac{3}{4}}
(\rho_\i+4\eps_\i)\eps^{\frac{11}{4}} -\eps^3\right]
^{-\frac{1}{2}}\right\}.\label{B:tau}
\end{eqnarray}
The corresponding asymptotic for pseudo-eccentricity is
\begin{eqnarray}
\e^{a-b}\sim\left[1+(\rho_\i+\eps_\i)/\l\right]^{\frac{1}{2}}
\left[1+F(0)\right]^{-1}.\label{B:ecc-asympt}
\end{eqnarray}
This form agrees with the general result expressed in
Eq.~(\ref{A:ab-asymptotic}). In addition, the lower bound for
pseudo-eccentricity can be derived: replacing the expression in the
braces in Eq.~(\ref{B:F}) by its bound,
$\l/\eps_\i+(1+\rho_\i/\eps_\i)x^{\frac{3}{4}}$, and integrating, we
find
\begin{eqnarray}
\e^{a-b}\gtrsim\left[\frac{\rho_\i}{\rho_\i+\eps_\i}
+\frac{\eps_\i}{\rho_\i+\eps_\i}\left(1+\frac{\rho_\i+\eps_\i}{\l}\right)
^{-\frac{1}{2}}\right]^{-1}\ge 1.\label{B:ineq}
\end{eqnarray}
Hence, the asymmetric expansion is always oblate in this case.

We finally note that in the case of zero vacuum energy, the exact
solution derived above simplifies significantly since integrals in
both Eqs.~(\ref{B:F}) and (\ref{B:integral0}) are elementary
functions:
\begin{eqnarray}
F(\eps/\eps_\i)&=&(\rho_\i/\eps_\i)\frac{(1+\rho_\i/\eps_\i)^{\frac{1}{2}}}
{(4+\rho_\i/\eps_\i)^3}
\left\{\frac{108-(10+\rho_\i/\eps_\i)^2}{(1+\rho_\i/\eps_\i)^\frac{1}{2}}\right.\nn\\
&&\hspace{20ex}\left.- \frac{108(\eps/\eps_\i)^\frac{1}{2}
-\left[4+\rho_\i/\eps_\i+6(\eps/\eps_\i)^\frac{1}{4}\right]^2}
{(\eps/\eps_\i)^{\frac{3}{8}}
\left[4+\rho_\i/\eps_\i-3(\eps/\eps_\i)^\frac{1}{4}\right]^\frac{1}{2}}\right\},
\label{B:F-exact}\\
t-t_\i&=&2(3\eps_\i)^{-\frac{1}{2}}
\left\{\frac{\left[4+\rho_\i/\eps_\i+6(\eps/\eps_\i)^\frac{1}{4}
\right]\left[4+\rho_\i/\eps_\i-3(\eps/\eps_\i)^\frac{1}{4}
\right]^\frac{1}{2}}{(\eps/\eps_\i)^\frac{3}{8}(4+\rho_\i/\eps_\i)^2}
\right.\nn\\ &&\hspace{11ex}\left.-
\frac{(10+\rho_\i/\eps_\i)(1+\rho_\i/\eps_\i)^\frac{1}{2}}
{(4+\rho_\i/\eps_\i)^2}\right\}.
\label{B:integral-exact}
\end{eqnarray}
The resulting form of the solution is then given by Eqs.~(\ref{B:a0}),
(\ref{B:rho0}), (\ref{B:b0}), (\ref{B:F-exact}),
(\ref{B:integral-exact}). Using this solution we find the following
large-time asymptotics:
\begin{eqnarray}
&&\eps\sim \eps_\i \left[\fr{1}{2}(4+\rho_\i/\eps_\i)^{\frac{1}{2}}
(3\eps_\i)^\frac{1}{2}t\right]^{-\frac{8}{3}},
\label{B:e0-asympt}\\ &&a\sim \fr{2}{3} 
\ln{\left[\fr{1}{2}(4+\rho_\i/\eps_\i)^{\frac{1}{2}}
(3\eps_\i)^\frac{1}{2}t\right]},
\label{B:a0-asympt}\\ &&b\sim \fr{2}{3} 
\ln{\left[\fr{1}{2}(\rho_\i/\eps_\i)^{\frac{3}{2}}
(4+\rho_\i/\eps_\i)^{-1} (3\eps_\i)^\frac{1}{2}(t-t_\i)\right]},
\label{B:b0-asympt}\\ && \rho\sim 
\fr{4}{3}t^{-2}.\label{B:rho0-asympt}
\end{eqnarray}
The pseudo-eccentricity becomes
\begin{eqnarray}
\e^{a-b}\sim 1+4\eps_\i/\rho_\i.\label{B:ecc0-asympt}
\end{eqnarray}


\begin{figure}
\includegraphics[width=10cm]{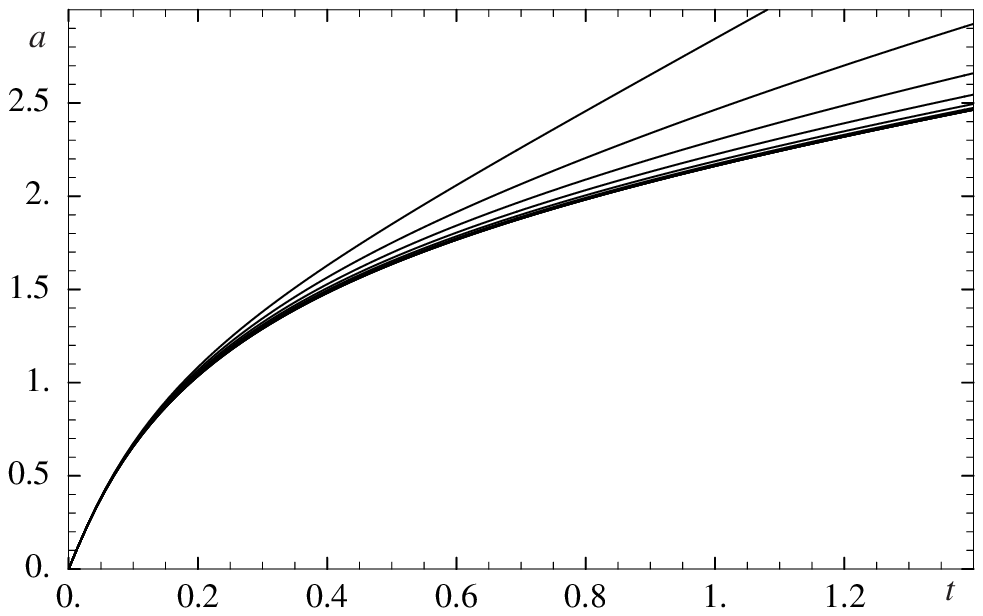}
\caption{Expansion parameter $a$ as a function of $t$ for the case
  $\text{M}\Lambda w$ with $\lambda=1$, $\rho_\text{i}=10$,
  $\epsilon_\text{i}=200$. Curves are for $w$ from $-1$ to $1$ with
  step $0.2$ from top to bottom.
  \label{fig-m-a}}
\end{figure}

\begin{figure}
\includegraphics[width=10cm]{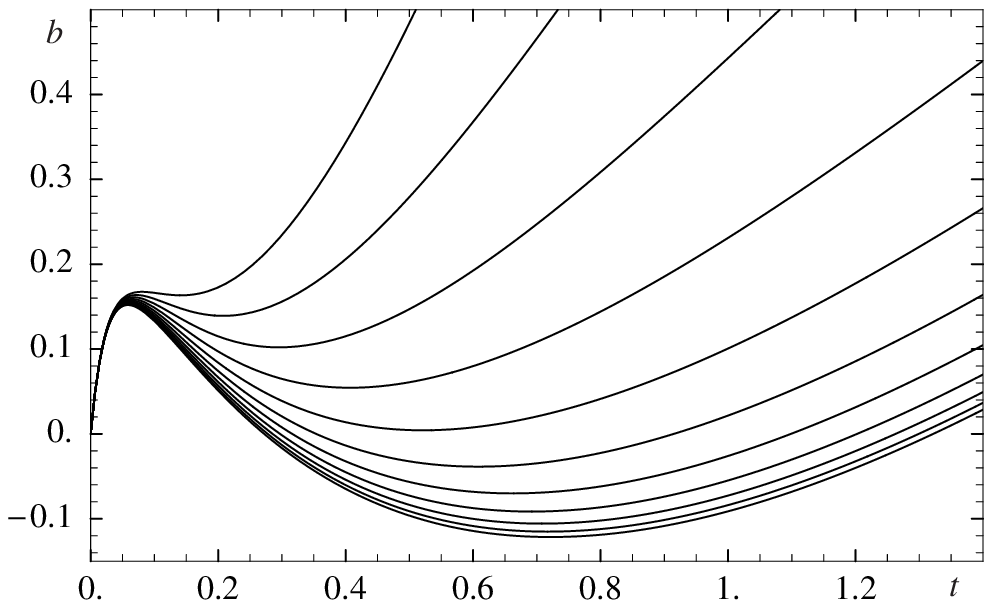}
\caption{Expansion parameter $b$ as a function of $t$ for the case
  $\text{M}\Lambda w$ with $\lambda=1$, $\rho_\text{i}=10$,
  $\epsilon_\text{i}=200$. Curves are for $w$ from $-1$ to $1$ with
  step $0.2$ from top to bottom.
  \label{fig-m-b}}
\end{figure}

\begin{figure}
\includegraphics[width=10cm]{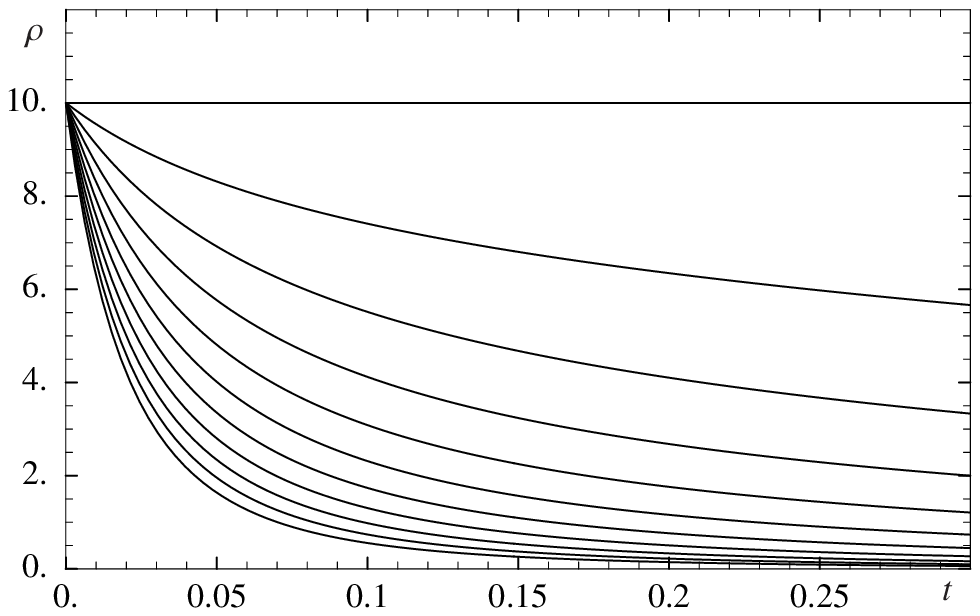}
\caption{Matter density $\rho$ as a function of $t$ for the case
  $\text{M}\Lambda w$ with $\lambda=1$, $\rho_\text{i}=10$,
  $\epsilon_\text{i}=200$. Curves are for $w$ from $-1$ to $1$ with
  step $0.2$ from top to bottom.
  \label{fig-m-r}}
\end{figure}

\begin{figure}
\includegraphics[width=10cm]{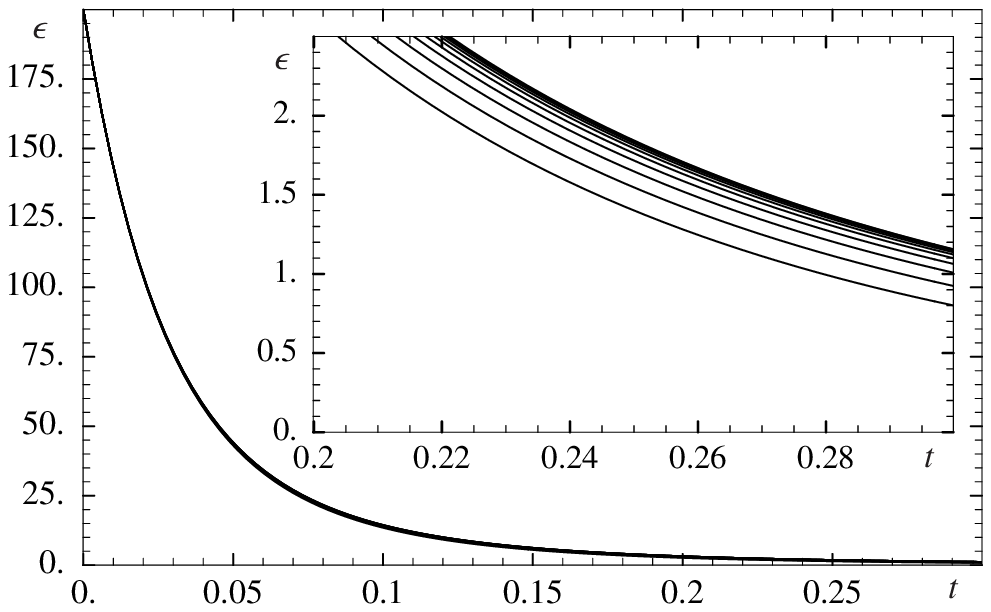}
\caption{Magnetic field density $\epsilon$ as a function of $t$ for
  the case $\text{M}\Lambda w$ with $\lambda=1$, $\rho_\text{i}=10$,
  $\epsilon_\text{i}=200$. Curves are for $w$ from $-1$ to $1$ with
  step $0.2$ from bottom to top.
  \label{fig-m-e}}
\end{figure}

\begin{figure}
\includegraphics[width=10cm]{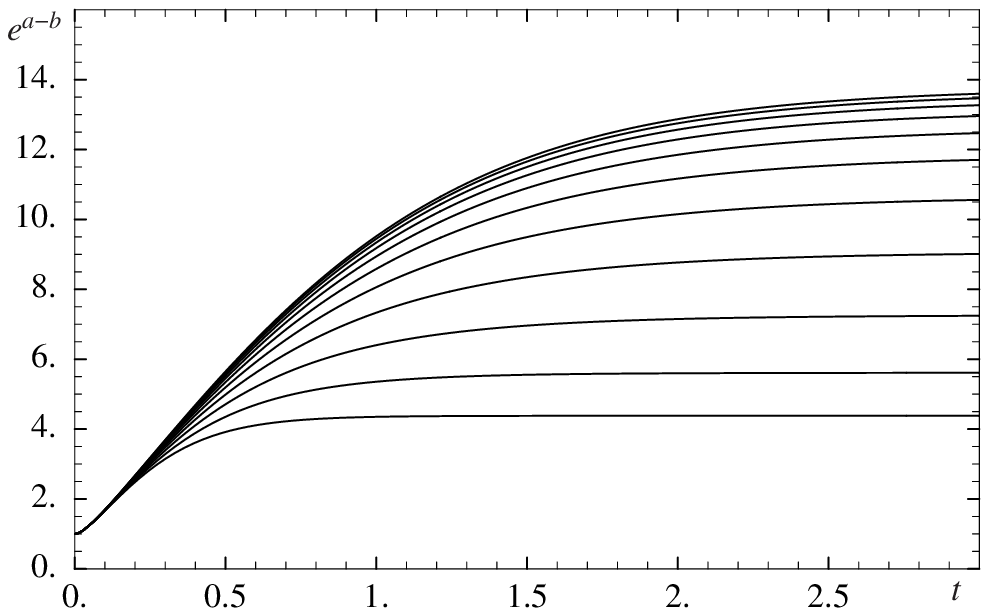}
\caption{Pseudo-eccentricity $e^{a-b}$ for the case $\text{M}\Lambda
  w$ with $\lambda=1$, $\rho_\text{i}=10$,
  $\epsilon_\text{i}=200$. Curves are for $w$ from $-1$ to $1$ with
  step $0.2$ from bottom to top. \label{fig-m-eccentricity-new}}
\end{figure}

\begin{figure}
\includegraphics[width=10cm]{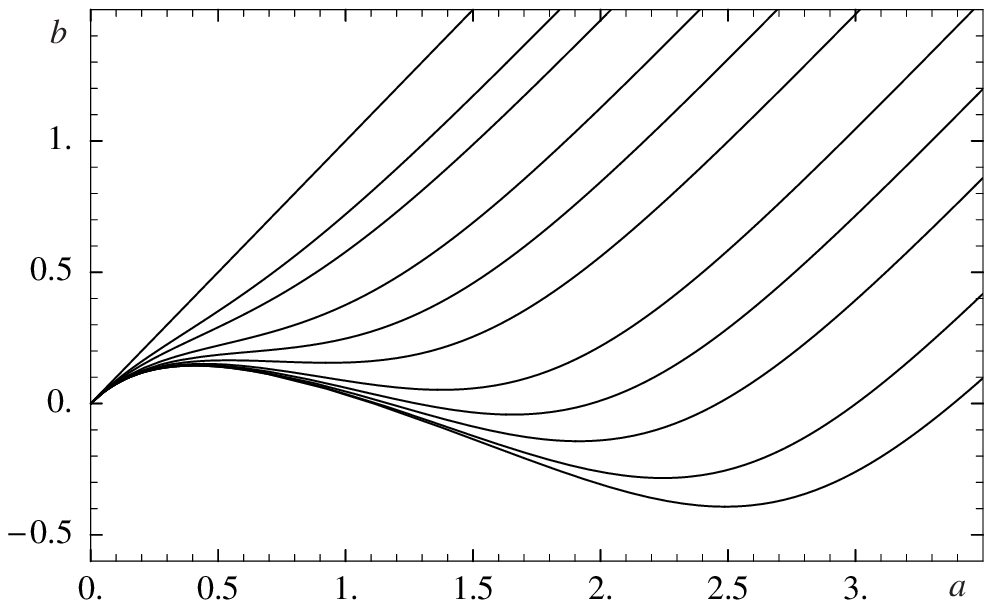}
\caption{Expansion parameters $a$ and $b$ for the case
  $\text{M}\Lambda$ with $\lambda=1$, $\rho_\text{i}=0$. Curves are
  for $\epsilon_\text{i}=0,1,2,5,10,20,50,100,200,500,1000$ from top
  to bottom. \label{fig-m-ab-new}}
\end{figure}

\begin{figure}
  \includegraphics[width=10cm]{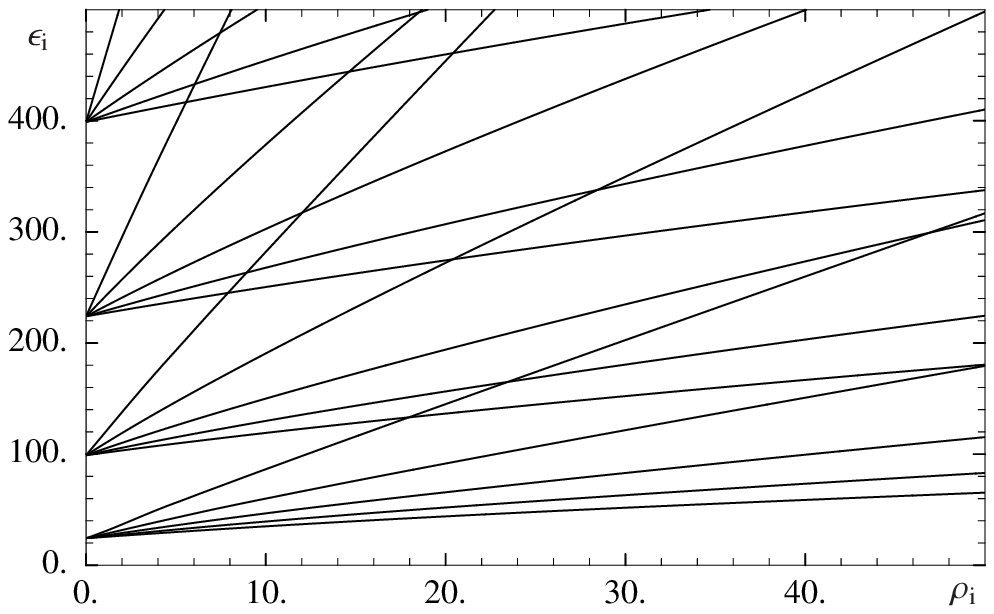}
  \caption{Asymptotic value of the pseudo-eccentricity for the case
    $\text{M}\Lambda w$ with $\lambda=1$ as a function of
    $\rho_\text{i}$ and $\epsilon_\text{i}$. Sets of curves are for
    $\e^{a-b}$ equal to $20,15,10,5$ from top to bottom; the abscissa
    corresponds to $\e^{a-b}=1$. Curves in each set are for $w$ equal
    to $-0.5, -0.25, 0, 0.25, 0.5$ from top to bottom.
    \label{fig-m-eccentricity-2}}
\end{figure}

\begin{figure}
  \includegraphics[width=10cm]{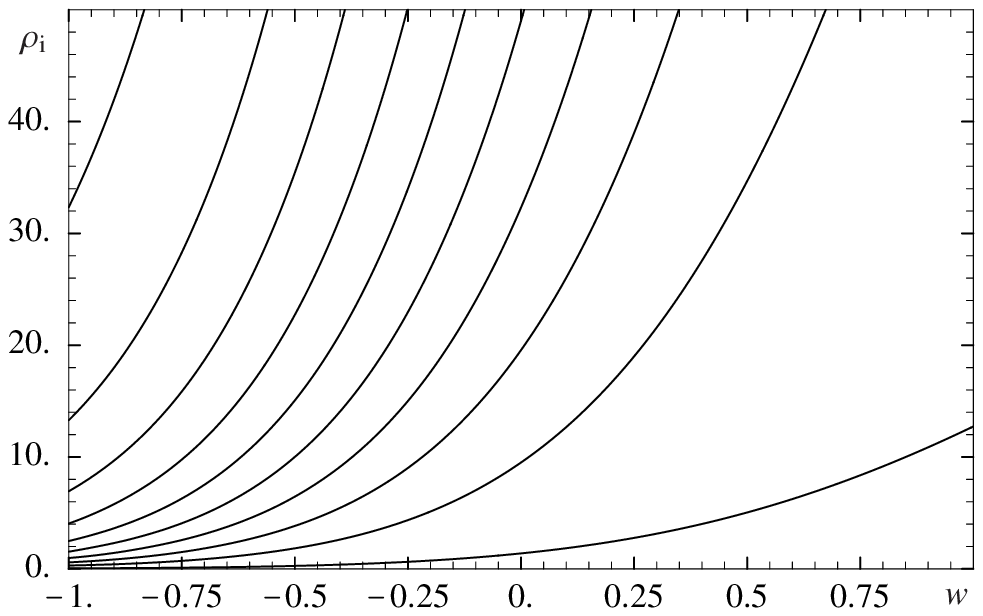}
  \caption{Asymptotic value of the pseudo-eccentricity for the case
    $\text{M}\Lambda w$ with $\lambda=1$ as a function of $w$ and
    $\rho_\text{i}$ for $\epsilon_\text{i}=200$. Curves are for
    $\e^{a-b}$ from $4$ to $22$ with step $2$ from top to
    bottom. \label{fig-m-eccentricity-2-new}}
\end{figure}


In the following, we will refer to the parameters $a$ and $b$ as
planar and axial expansion parameters.  Note that in order to generate
the figures, we have used large values for the initial matter density,
$\rho_\i$ and the magnetic field energy density $\epsilon_\i$ relative
to the cosmological constant $\lambda$ in order to make the effects of
their contributions to the stress energy tensor stand out in the
graphics. We will do this throughout the paper, but caution the reader
that we are not implying these are realistic choices of parameters. A
realistic choice of initial conditions would probably be $\lambda$,
$\rho_\i$, and $\epsilon_\i$ all of the same order of magnitude, but
in this case we would need to plot small differences of parameters
instead of plotting them directly. Recall that the $\delta\rho/\rho$
effects found in the cosmic microwave background density perturbations
are of order $10^{-5}$, so observationally one is typically, but not
always, looking for small effects.

Figure 1 gives the expansion parameter $a$ as a function of time in a
universe filled with aligned magnetic fields, cosmological constant,
and matter. Each curve corresponds to matter with a different equation
of state parameter $w$.

Each curve in Figure 2 plots the axial expansion parameter $b$ with
time in a universe filled with aligned magnetic fields and
cosmological constant for matter with a variety of choices for
$w$. Comparing Figs. 1 and 2 shows $a$ grows faster than $b$. This
implies the expansion is oblate, i.~e., an initial spherical region
expands to an oblate spheroid.

Figures 3 and 4 show the decay of the matter density and of the
magnetic field energy respectively, with time using the same initial
parameters that were used to generate Figs.~1 and 2. Figure 5 shows
the behavior of the pseudo-eccentricity with time, again for the same
initial parameters which were used in the previous figures.

In Fig. 6 we have plotted the axial expansion parameter versus the
planar expansion parameter. Each curve is for a different value of
initial magnetic field energy density. Time increases along each curve
from left to right. Note that $a$ always increases, but for
sufficiently strong initial magnetic fields, after an initial
increase, $b$ reaches a maximum, then decreases for a time, reaches a
minimum, and then increases thereafter. Figure 7 shows the asymptotic
values of the pseudo-eccentricity as a function of magnetic field
energy density and initial matter density for various fixed values of
$w$. As expected, stronger $\epsilon_\i$ leads to higher eccentricity,
but increasing $\rho_\i$, the spherically symmetric component of
${T^{\mu}}_{\nu}$ tends to dampen the effect. As with previous
figures, the parameters in Figs.~6 and 7 were chosen to enhance the
visualization, not for physical reasons. Finally, Fig.~8 is a contour
plot of the asymptotic value of the pseudo-eccentricity (the curves
are lines of equal asymptotic value of $e^{a-b}$) for a range of
initial matter densities and equations of state.

\section{Strings}\label{S:S}

In a somewhat artificial case of cosmological constant plus strings
plus matter, the equations to solve are
\begin{eqnarray}
&&\dot{a}^2+2\dot{a}\dot{b}=\l+\rho+\eps,\label{S:1}\\
&&2\ddot{a}+3\dot{a}^2=\l-w\rho+\eps,\label{S:3}\\
&&\dot{\eps}+2\dot{a}\eps=0.\label{S:4}
\end{eqnarray}
From the conservation of the anisotropic part of the energy-momentum,
Eq.~(\ref{S:4}), we find
\begin{eqnarray}
a=\fr{1}{2}\ln{(\eps_\i/\eps)}.\label{S:a0}
\end{eqnarray}
Proceeding with the analysis in a matter similar to Sec.~\ref{S:B}, we
arrive at the following system of equations:
\begin{eqnarray}
&&\eps f'-\fr{7}{2}f+\eps^2(\l-w\rho+\eps)=0,\label{S:f-eq}\\ &&\eps
f\rho'-(1+w)\rho\left[\fr{3}{4}f
+\fr{1}{2}\eps^2(\l+\rho+\eps)\right]=0,\label{S:rho-eq}
\end{eqnarray}
or equivalently
\begin{eqnarray}
&&f=\fr{2}{3}\l\eps^2+2\eps^3+\fr{2}{3}\eps_\i^{-\frac{3}{2}}(\rho_\i-2\eps_\i)
\eps^{\frac{7}{2}}-w\eps^{\frac{7}{2}}
\int_\eps^{\eps_\i}\d\eps\,\rho\eps^{-\frac{5}{2}},\label{S:f}\\
&&\rho=\rho_\i(\eps/\eps_\i)^{\frac{3}{4}(1+w)}\psi
\left[1+\fr{1}{2}(1+w)\rho_\i\eps_\i\int_\eps^{\eps_\i}\d\eps\,
(\eps/\eps_\i)^{1+\frac{3}{4}(1+w)}\psi
f^{-1}\right]^{-1},\label{S:rho}
\end{eqnarray}
where
\begin{equation} 
\psi=\exp{\left[-\fr{1}{2}(1+w)
\int_\eps^{\eps_\i}\d\eps\,\eps(\l+\eps)f^{-1}\right]}.\label{S:psi}
\end{equation} 

As in the magnetic field case, we can find the exact solution to the
coupled system of equations only for non-relativistic matter ($w=0$),
which we present below. An approximate solution for arbitrary $w$
together with large $t$ asymptotics are given in
Appendix~\ref{S:S0}. Setting $w=0$, Eqs.~(\ref{S:f}) and (\ref{S:rho})
reduce to
\begin{eqnarray}
&&f=\fr{2}{3}\l\eps^2+2\eps^3+\fr{2}{3}\eps_\i^{-\frac{3}{2}}(\rho_\i-2\eps_\i)
\eps^{\frac{7}{2}},\label{S:f0}\\
&&\rho=\rho_\i(\eps/\eps_\i)^\frac{3}{2}\left[1+F(\eps/\eps_\i)\right]^{-1}
\left[\frac{\l+3\eps+(\rho_\i-2\eps_\i)(\eps/\eps_\i)^\frac{3}{2}}
{\l+\rho_\i+\eps_\i}\right]^{-\frac{1}{2}},\label{S:rho0}
\end{eqnarray}
where
\begin{eqnarray}
F(\eps/\eps_\i)
=\fr{3}{4}(\rho_\i/\eps_\i)[1+(\l+\rho_\i)/\eps_\i]^{\frac{1}{2}}
\int_{\eps/\eps_\i}^{1}\d x\,x^{\frac{1}{2}}
\left\{\l/\eps_\i+3x+(\rho_\i/\eps_\i-2)x^{\frac{3}{2}}\right\}^{-\frac{3}{2}}.
\label{S:F}
\end{eqnarray}
Equations~(\ref{A5s}), (\ref{S:a0}) and (\ref{S:rho0}) give
\begin{eqnarray}
b=\fr{1}{2}\ln\frac{\l+3\eps+(\rho_\i-2\eps_\i)(\eps/\eps_\i)^{\frac{3}{2}}}
{\l+\rho_\i+\eps_\i}-\fr{1}{2}\ln{(\eps/\eps_\i)}
+\ln\left[1+F(\eps/\eps_\i)\right].
\label{S:b0}
\end{eqnarray}
The time dependences of the above functions $\rho(\eps)$, $a(\eps)$
and $b(\eps)$ are found from the function $\eps(t)$ which is given
implicitly by
\begin{eqnarray}
t-t_\i=\fr{1}{2} \int^{\eps_\i}_{\eps}\d\eps\left[
\fr{1}{3}\l\eps^2+\eps^3+\fr{1}{3}\eps_\i^{-\frac{3}{2}}(\rho_\i-2\eps_\i)
\eps^{\frac{7}{2}}\right]^{-\frac{1}{2}}.\label{S:integral0}
\end{eqnarray}

There is a substantial analogy with the magnetic field case: all the
statements in the paragraph following Eq.~(\ref{B:integral0}) in
Sec.~\ref{S:B} are also correct for the case of strings if one
replaces the magnetic field density with the string density.

In the case $w=0$, asymptotics from Appendix~\ref{S:S0} simplify as
follows:
\begin{eqnarray}
&&\eps\sim
\eps_\i\,\exp\left[-2(\l/3)^{\frac{1}{2}}(t-t_\i+\tau)\right],
\label{S:e-asympt}\\ &&a\sim(\l/3)^{\frac{1}{2}}(t-t_\i+\tau), 
\label{S:a-asympt}\\ &&b\sim (\l/3)^{\frac{1}{2}}(t-t_\i+\tau)-\fr{1}{2}
\ln{\left[1+(\rho_\i+\eps_\i)/\l\right]}+\ln{[1+F(0)]},
\label{S:b-asympt}\\ && \rho\sim\rho_\i 
\left[1+(\l+\rho_\i)/\eps_\i\right]^{\frac{1}{2}}[1+F(0)]^{-1}
\exp\left[-(3\l)^{\frac{1}{2}}(t-t_\i+\tau)\right],\label{S:rho-asympt}
\end{eqnarray}
where
\begin{eqnarray}
\tau=\fr{1}{2}\int_{0}^{\eps_\i}
\d\eps\left\{\left(\fr{1}{3}\l\eps^2\right)^{-\frac{1}{2}}
-\left[\fr{1}{3}\l\eps^2+\eps^3+\fr{1}{3}\eps_\i^{-\frac{3}{2}}(\rho_\i-2\eps_\i)
\eps^{\frac{7}{2}}\right]^{-\frac{1}{2}}\right\}.\label{S:tau}
\end{eqnarray}
The corresponding asymptotic for the pseudo-eccentricity is
\begin{eqnarray}
\e^{a-b}\sim\left[1+(\rho_\i+\eps_\i)/\l\right]^{\frac{1}{2}}
\left[1+F(0)\right]^{-1}.\label{S:ecc-asympt}
\end{eqnarray}
To find the lower bound for this quantity, we replace the expression
in the braces in Eq.~(\ref{S:F}) by its bound,
$\l/\eps_\i+(1+\rho_\i/\eps_\i)x^{\frac{3}{2}}$, integrate, and find
\begin{eqnarray}
\e^{a-b}\gtrsim\left[\frac{\rho_\i}{\rho_\i+\eps_\i}
+\frac{\eps_\i}{\rho_\i+\eps_\i}\left(1+\frac{\rho_\i+\eps_\i}{\l}\right)
^{-\frac{1}{2}}\right]^{-1}\ge 1.\label{S:ineq}
\end{eqnarray}
This is the same bound we found for the magnetic field case
[Eq.~(\ref{B:ineq})], and we see the expansion is again of the oblate
form.

Again, for the case $\l=0$, the above solution is given in terms of
elementary functions. We find
\begin{eqnarray}
F(\eps/\eps_\i)&=&(\rho_\i/\eps_\i)\theta(1)
\left\{\frac{1}{\theta(1)}-\frac{1}{\theta(\eps/\eps_\i)}
-\arctanh{\theta(1)}+\arctanh{\theta(\eps/\eps_\i)} \right\}\\
t-t_\i&=&\eps_\i^{-\frac{1}{2}}\left\{\fr{1}{3}(\rho_\i/\eps_\i-2)
\left[\arctanh{\theta(1)}-\arctanh{\theta(\eps/\eps_\i)}\right]\right.\nn\\
&&\hspace{5ex}\left.-\theta(1)
+(\eps/\eps_\i)^{-\frac{1}{2}}\theta(\eps/\eps_\i) \right\},
\end{eqnarray}
where 
\begin{eqnarray}
\theta(x)=\left[1+\fr{1}{3}\left(\rho_\i/\eps_\i-2\right)
x^{\frac{1}{2}}\right]^{\frac{1}{2}}.
\end{eqnarray}

The large-time asymptotics are 
\begin{eqnarray}
&&\eps\sim\eps_\i f^{-2}(t),\label{S:e0-asympt}\\ 
&&a\sim f(t),\label{S:a0-asympt}\\ 
&&b\sim -f(t),\label{S:b0-asympt}\\ 
&&\rho\sim\rho_\i\e^{-f(t)},\label{S:rho0-asympt}
\end{eqnarray}
where $f(t)=t\eps_\i^{\frac{1}{2}}+\fr{1}{6}(\rho_\i/\eps_\i-2)
\ln{(t\eps_\i^{\frac{1}{2}})}$. The pseudo-eccentricity becomes
\begin{eqnarray}
\e^{a-b}\sim \e^{2f(t)}.\label{S:ecc0-asympt}
\end{eqnarray}


\begin{figure}
\includegraphics[width=10cm]{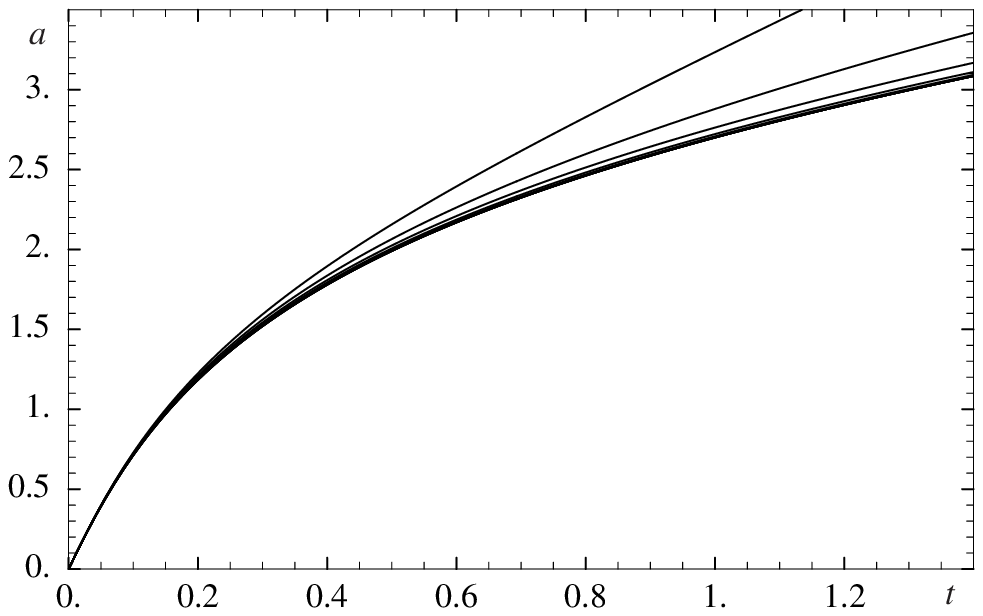}
\caption{Expansion parameter $a$ as a function of $t$ for the case
  $\text{S}\Lambda w$ with $\lambda=1$, $\rho_\text{i}=10$,
  $\epsilon_\text{i}=200$. Curves are for $w$ from $-1$ to $1$ with
  step $0.2$ from top to bottom.
  \label{fig-s-a}}
\end{figure}

\begin{figure}
\includegraphics[width=10cm]{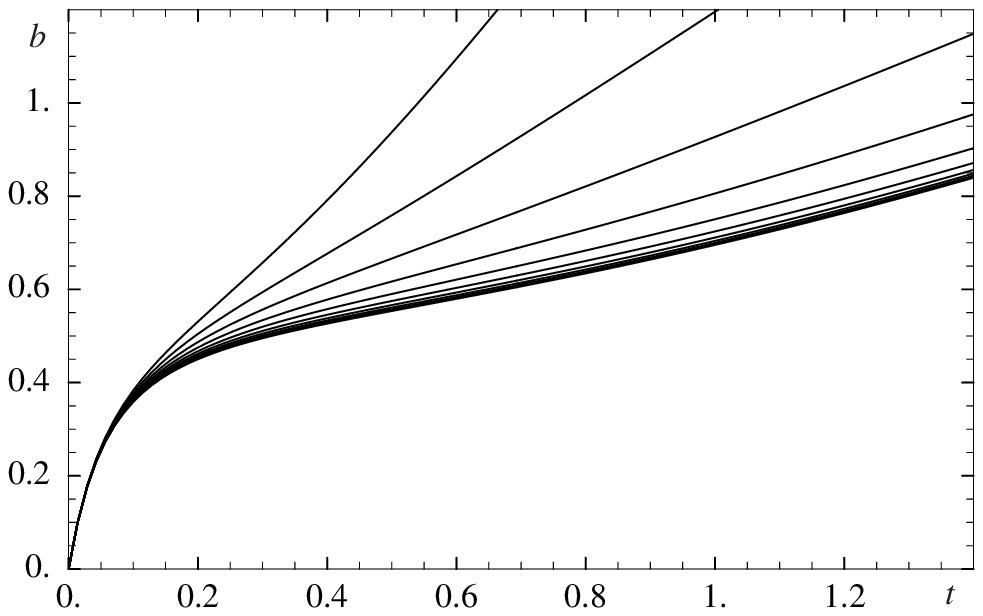}
\caption{Expansion parameter $b$ as a function of $t$ for the case
  $\text{S}\Lambda w$ with $\lambda=1$, $\rho_\text{i}=10$,
  $\epsilon_\text{i}=200$. Curves are for $w$ from $-1$ to $1$ with
  step $0.2$ from top to bottom.
  \label{fig-s-b}}
\end{figure}

\begin{figure}
\includegraphics[width=10cm]{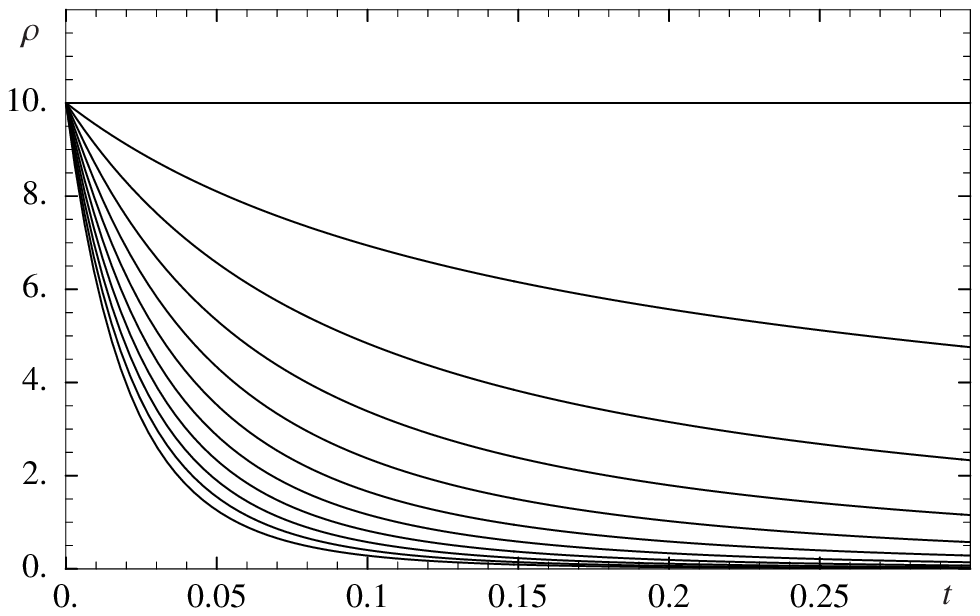}
\caption{Matter density $\rho$ as a function of $t$ for the case
  $\text{S}\Lambda w$ with $\lambda=1$, $\rho_\text{i}=10$,
  $\epsilon_\text{i}=200$. Curves are for $w$ from $-1$ to $1$ with
  step $0.2$ from top to bottom.
  \label{fig-s-r}}
\end{figure}

\begin{figure}
\includegraphics[width=10cm]{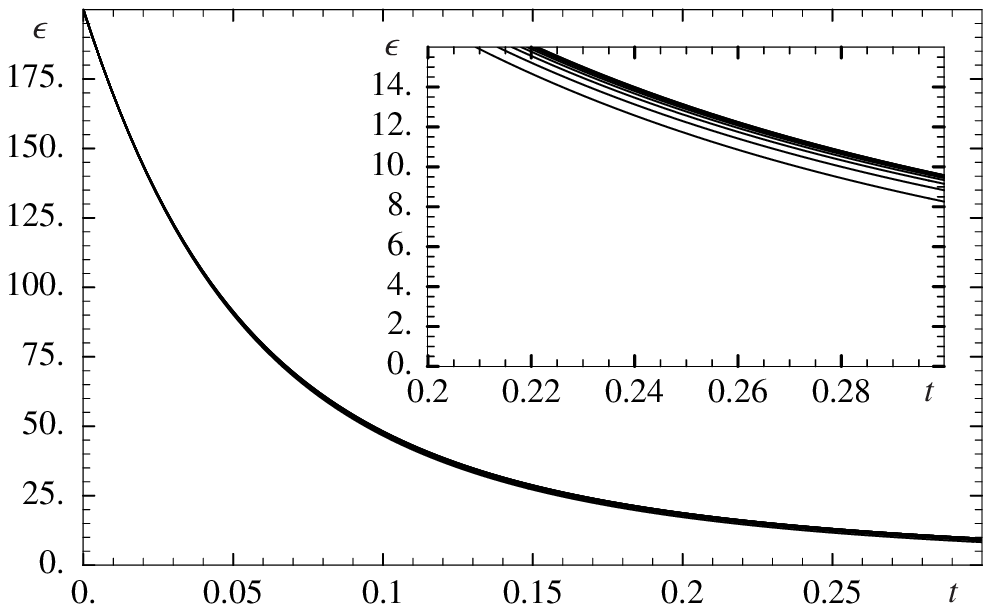}
\caption{Magnetic field density $\epsilon$ as a function of $t$ for
  the case $\text{S}\Lambda w$ with $\lambda=1$, $\rho_\text{i}=10$,
  $\epsilon_\text{i}=200$. Curves are for $w$ from $-1$ to $1$ with
  step $0.2$ from bottom to top.
  \label{fig-s-e}}
\end{figure}

\begin{figure}
\includegraphics[width=10cm]{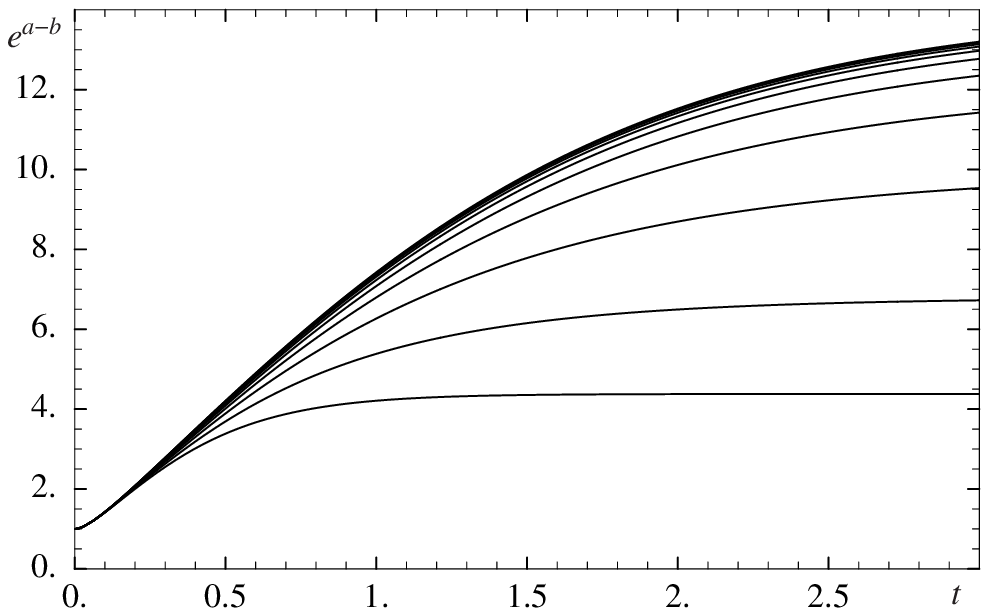}
\caption{Pseudo-eccentricity $e^{a-b}$ for the case $\text{S}\Lambda
  w$ with $\lambda=1$, $\rho_\text{i}=10$,
  $\epsilon_\text{i}=200$. Curves are for $w$ from $-1$ to $1$ with
  step $0.2$ from bottom to top. \label{fig-s-eccentricity-1}}
\end{figure}

\begin{figure}
\includegraphics[width=10cm]{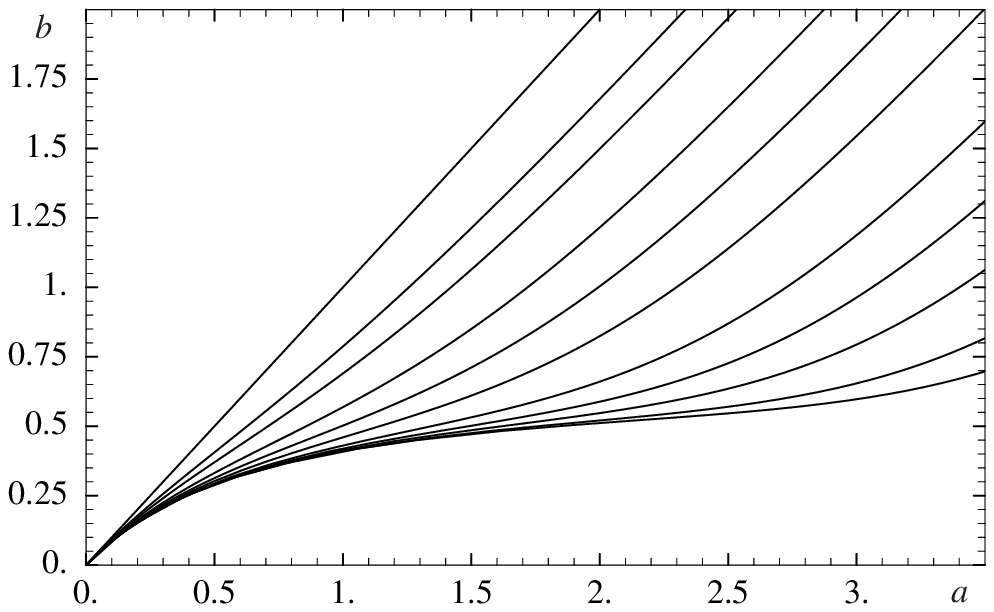}
\caption{Expansion parameters $a$ and $b$ for the case
  $\text{S}\Lambda$ with $\lambda=1$, $\rho_\text{i}=0$. Curves are
  for $\epsilon_\text{i}=0,1,2,5,10,20,50,100,200,500,1000$ from top
  to bottom. \label{fig-s-ab}}
\end{figure}

\begin{figure}
  \includegraphics[width=10cm]{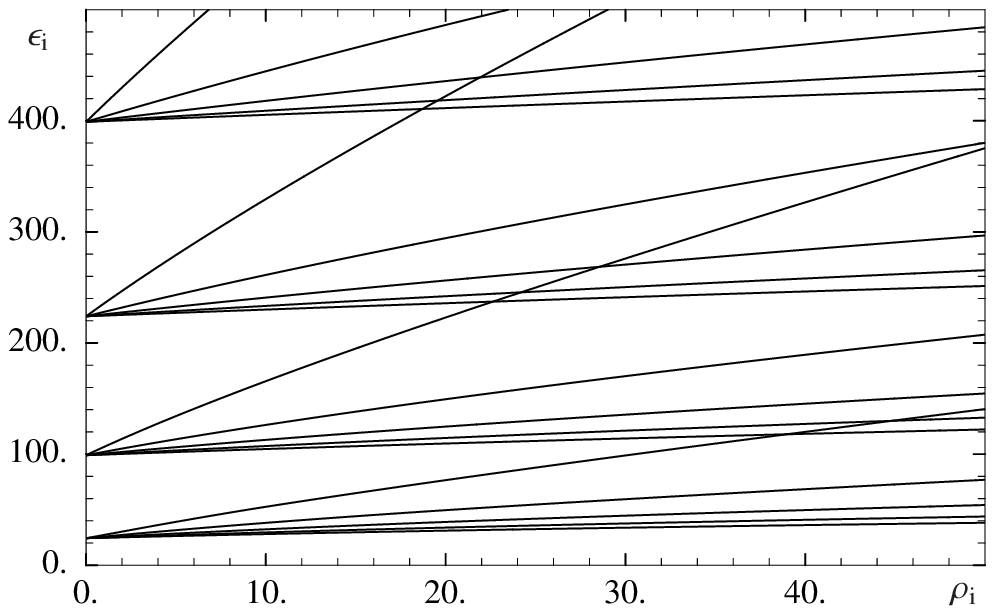}
  \caption{Asymptotic value of the pseudo-eccentricity for the case
    $\text{S}\Lambda w$ with $\lambda=1$ as a function of
    $\rho_\text{i}$ and $\epsilon_\text{i}$. Sets of curves are for
    $\e^{a-b}$ equal to $20,15,10,5$ from top to bottom; the abscissa
    corresponds to $\e^{a-b}=1$. Curves in each set are for $w$ equal
    to $-0.5, -0.25, 0, 0.25, 0.5$ from top to bottom.
    \label{fig-s-eccentricity-2}}
\end{figure}

\begin{figure}
  \includegraphics[width=10cm]{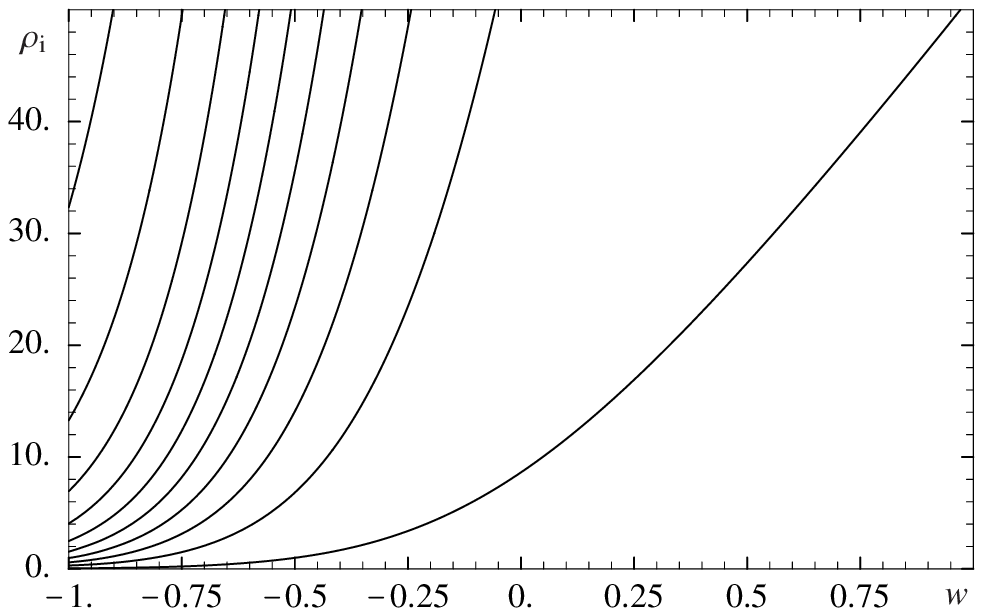}
  \caption{Asymptotic value of the pseudo-eccentricity for the case
    $\text{S}\Lambda w$ with $\lambda=1$ as a function of $w$ and
    $\rho_\text{i}$ for $\epsilon_\text{i}=200$. Curves are for
    $\e^{a-b}$ from $4$ to $22$ with step $2$ from top to
    bottom. \label{fig-s-eccentricity-3}}
\end{figure}


Figures~9 and 10 plot $a$ and $b$ as a function of time for a range of
$w$ values. While Fig.~9 is qualitatively similar to Fig.~1, Fig.~10
shows only a monotonic increase in $b$, unlike Fig. 2. (See {\bf
\ref{i:dot-a}}, {\bf \ref{i:dot-b}} in Sec.~\ref{S:G}.) Figures~11 and
12 show the matter density and string density as a function of
time. These figures are qualitatively similar to Figs.~3 and 4 for the
magnetic field case, but the curves in the string case are somewhat
more compressed. Figure~13 is the plot of the pseudo-eccentricity with
time for strings. Figure~14 plots $a$ versus $b$ for a variety of
values of initial string density. The monotonic increase of $b$ is
again apparent, in contrast to the results for magnetic fields shown
in Fig.~6.

Figure~15 gives contours of asymptotic values of the
pseudo-eccentricity as a function of matter density and string
density. The results are similar, but somewhat milder than the
magnetic case, Fig.~7. Finally, Fig.~16. shows a dependence of the
asymptotic value of the pseudo-eccentricity on the equation of
state. The effect is again similar to, but milder than, the magnetic
field case.

\section{Walls}\label{S:W}

If we replace magnetic fields or cosmic strings from the previous two
sections by a uniform stack of cosmic domain walls, we arrive at
another solvable model described by the following equations:
\begin{eqnarray}
&&\dot{a}^2+2\dot{a}\dot{b}=\l+\rho+\eps,\label{W:1}\\
&&2\ddot{a}+3\dot{a}^2=\l-w\rho,\label{W:3}\\
&&\dot{\eps}+\dot{b}\eps=0.\label{W:4}
\end{eqnarray}
Again, we were not able to solve the above equations exactly for
arbitrary $w$; also, it appears to be much harder to arrive at a
simple and accurate approximation similar to the approximations for
the cases of magnetic fields and strings (see Appendices~\ref{S:B0}
and \ref{S:S0} for details). We reluctantly restrict ourselves only to
the case $w=0$. Somewhat surprisingly, the analysis will need to be
substantially different from the magnetic fields and strings cases.

To proceed, it is convenient to use the substitution
$a=\frac{2}{3}\ln{u}$, which transforms the Riccati Eq.~(\ref{W:3})
into an easily solvable linear equation $\ddot{u}=\fr{3}{4}\l u$. This
results in
\begin{eqnarray}
a=\fr{2}{3}\ln\frac{\g^3-\s}{1-\s}-\ln{\g},\label{W:a}
\end{eqnarray}
where $\g=\exp\left[(\l/3)^{\frac{1}{2}}(t-t_\i)\right]$ and
\begin{eqnarray}
\s=\frac{\left[1+(\r_\i+\eps_\i)/\l\right]^{\frac{1}{2}}-1}
{\left[1+(\r_\i+\eps_\i)/\l\right]^{\frac{1}{2}}+1}.\label{W:sigma}
\end{eqnarray}
Energy-momentum conservation, Eq.~(\ref{W:4}), gives
\begin{eqnarray}
b=\ln{(\eps_\i/\eps)}.\label{W:b}
\end{eqnarray}
Using Eqs.~(\ref{A5s}), (\ref{W:1}), (\ref{W:a}), and (\ref{W:b}), we
find
\begin{eqnarray}
\eps=\eps_\i\g\frac{1+\s}{\g^3+\s}\left(\frac{\g^3-\s}{1-\s}\right)^{\frac{1}{3}}
\left\{1+\frac{3\eps_\i}{2\l}\frac{1+\s}{(1-\s)^{\frac{1}{3}}}
F(\g)+\frac{\r_\i}{2\l}(1-\s)\frac{\g^3-1}{\g^3+\s}\right\}^{-1},\label{W:e}
\end{eqnarray}
where 
\begin{eqnarray}
F(\g)=\int_1^\g \d x\,\frac{(x^3-\s)^{\frac{4}{3}}}{(x^3+\s)^2}.
\end{eqnarray}
The function $F(\g)$ can be written in terms of the hypergeometric
functions, but the expression is complicated and not illuminating, so
we will not give it here.

In the case $\l=0$, the above exact solution simplifies
significantly. Instead of Eq.~(\ref{W:a}) we now have
\begin{eqnarray}
a=\fr{2}{3}\ln(1-C+Ct),\label{W:a0}
\end{eqnarray}
where $C=1-\fr{1}{2}[3(\rho_\i+\eps_\i)]^{\frac{1}{2}}$, and
Eq.~(\ref{W:e}) becomes
\begin{eqnarray}
\eps=\eps_\i(1-C+Ct)^{\frac{1}{3}}\left\{1+\frac{3\rho_\i}{4C}(t-t_\i)
+\frac{9\eps_\i}{28C^2}\left[(1-C+Ct)^{\frac{7}{3}}-1\right]\right\}^{-1},
\label{W:e0}
\end{eqnarray}

Returning to the case of $\lambda\not=0$, the large $t$ wall energy
density is
\begin{eqnarray}
\eps\sim \eps_\i\,\e^{-(\l/3)^{\frac{1}{2}}
(t-t_\i)}\left[\frac{(1-\s)^{\frac{1}{3}}}{1+\s}
+\frac{3\eps_\i}{2\l}F(\infty)+\frac{\r_\i}{2\l}\frac{(1-\s)^{\frac{4}{3}}}{1+\s}
\right]^{-1},\label{W:e-asympt}
\end{eqnarray}
the transverse scale factor is
\begin{eqnarray}
a\sim(\l/3)^{\frac{1}{2}}(t-t_\i)
-\fr{2}{3}\ln{(1-\s)},\label{W:a-asympt}
\end{eqnarray}
the longitudinal scale factor is
\begin{eqnarray}
b\sim(\l/3)^{\frac{1}{2}}(t-t_\i)
+\ln\left[\frac{(1-\s)^{\frac{1}{3}}}{1+\s}
+\frac{3\eps_\i}{2\l}F(\infty)+\frac{\r_\i}{2\l}\frac{(1-\s)^{\frac{4}{3}}}{1+\s}
\right],\label{W:b-asympt}
\end{eqnarray}
and the matter density is
\begin{eqnarray}
\rho\sim
\e^{-(3\l)^{\frac{1}{2}}(t-t_\i)}\left[\frac{(1-\s)^{\frac{5}{3}}}{1+\s}
+\frac{3\eps_\i}{2\l}(1-\s)^{\frac{4}{3}}
F(\infty)+\frac{\r_\i}{2\l}\frac{(1-\s)^{\frac{8}{3}}}{1+\s}\right]^{-1}.
\end{eqnarray}
From Eqs.~(\ref{W:a-asympt}) and (\ref{W:b-asympt}) the
pseudo-eccentricity is
\begin{eqnarray}
\e^{a-b}\sim\left[\frac{1-\s}{1+\s}
+\frac{3\eps_\i}{2\l}(1-\s)^{\frac{2}{3}}
F(\infty)+\frac{\r_\i}{2\l}\frac{(1-\s)^2}{1+\s}\right]^{-1}.
\end{eqnarray}

In the case of zero vacuum energy, the asymptotics read:
\begin{eqnarray}
&&\eps\sim\text{const}\ t^{-2},\\
&&a\sim\fr{2}{3}\ln{t}+\text{const},\\ &&b\sim
2\ln{t}+\text{const},\\ &&\rho\sim\text{const}\ t^{-\frac{10}{3}}.
\end{eqnarray}
As opposed to the cases of magnetic fields and strings, the
pseudo-eccentricity vanishes for large $t$, since $\e^{a-b}\sim
t^{-\frac{4}{3}}$.


\begin{figure}
\includegraphics[width=10cm]{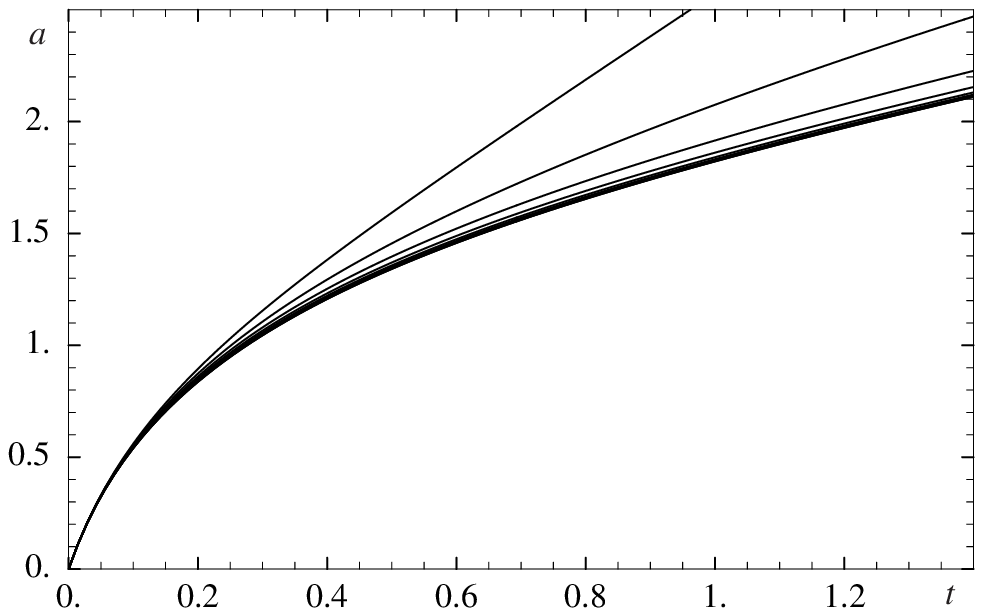}
\caption{Expansion parameter $a$ as a function of $t$ for the case
  $\text{W}\Lambda w$ with $\lambda=1$, $\rho_\text{i}=10$,
  $\epsilon_\text{i}=200$. Curves are for $w$ from $-1$ to $1$ with
  step $0.2$ from top to bottom.
  \label{fig-w-a}}
\end{figure}

\begin{figure}
\includegraphics[width=10cm]{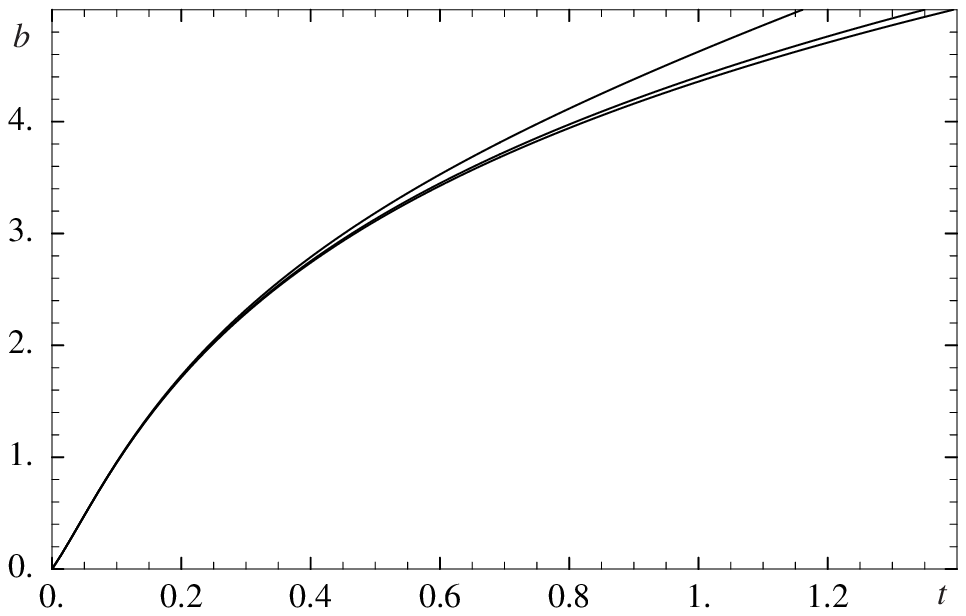}
\caption{Expansion parameter $b$ as a function of $t$ for the case
  $\text{W}\Lambda w$ with $\lambda=1$, $\rho_\text{i}=10$,
  $\epsilon_\text{i}=200$. Curves are for $w$ from $-1$ to $1$ with
  step $0.2$ from top to bottom.
  \label{fig-w-b}}
\end{figure}

\begin{figure}
\includegraphics[width=10cm]{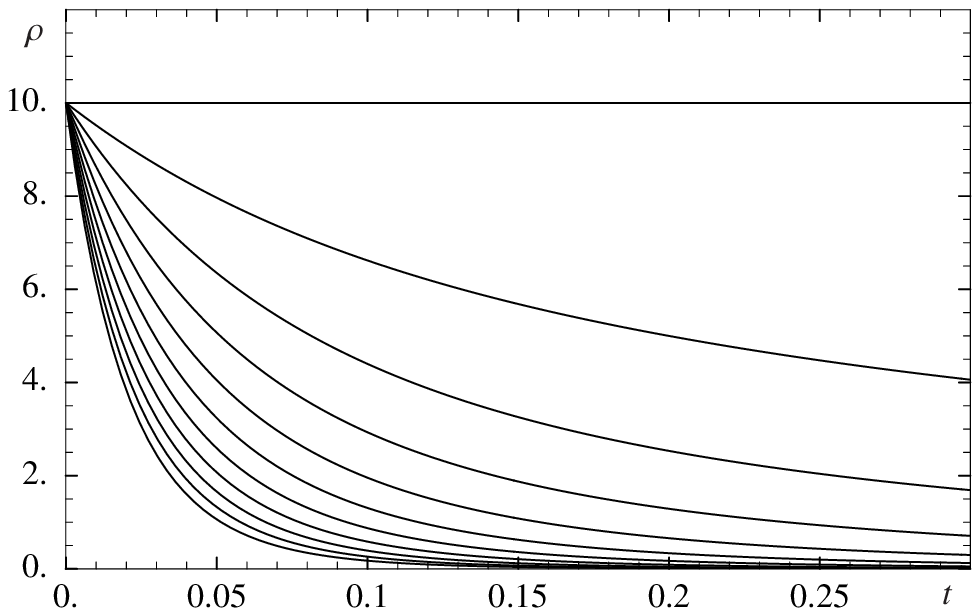}
\caption{Matter density $\rho$ as a function of $t$ for the case
  $\text{W}\Lambda w$ with $\lambda=1$, $\rho_\text{i}=10$,
  $\epsilon_\text{i}=200$. Curves are for $w$ from $-1$ to $1$ with
  step $0.2$ from top to bottom.
  \label{fig-w-r}}
\end{figure}

\begin{figure}
\includegraphics[width=10cm]{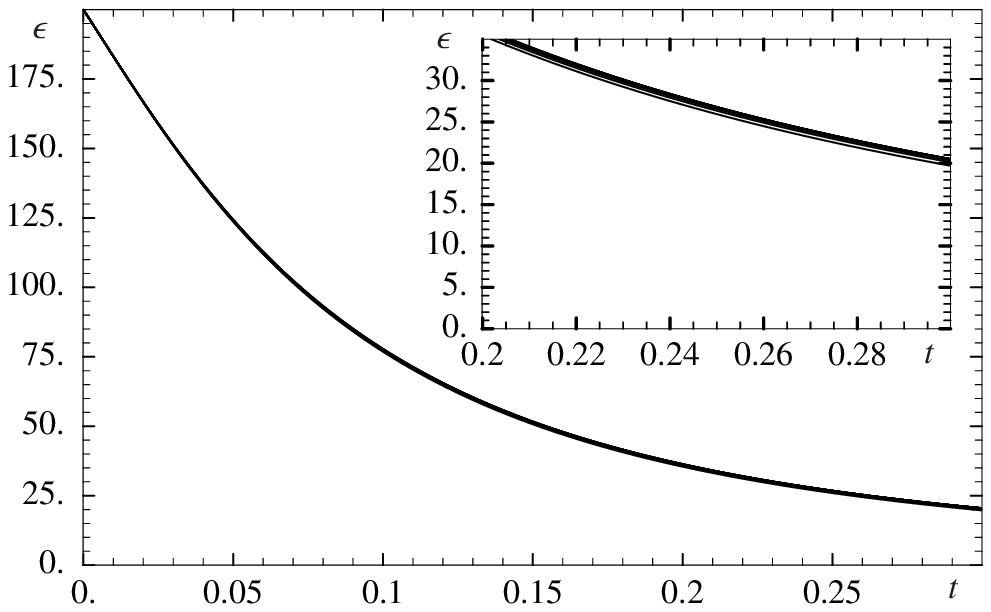}
\caption{Magnetic field density $\epsilon$ as a function of $t$ for
  the case $\text{W}\Lambda w$ with $\lambda=1$, $\rho_\text{i}=10$,
  $\epsilon_\text{i}=200$. Curves are for $w$ from $-1$ to $1$ with
  step $0.2$ from bottom to top.
  \label{fig-w-e}}
\end{figure}

\begin{figure}
\includegraphics[width=10cm]{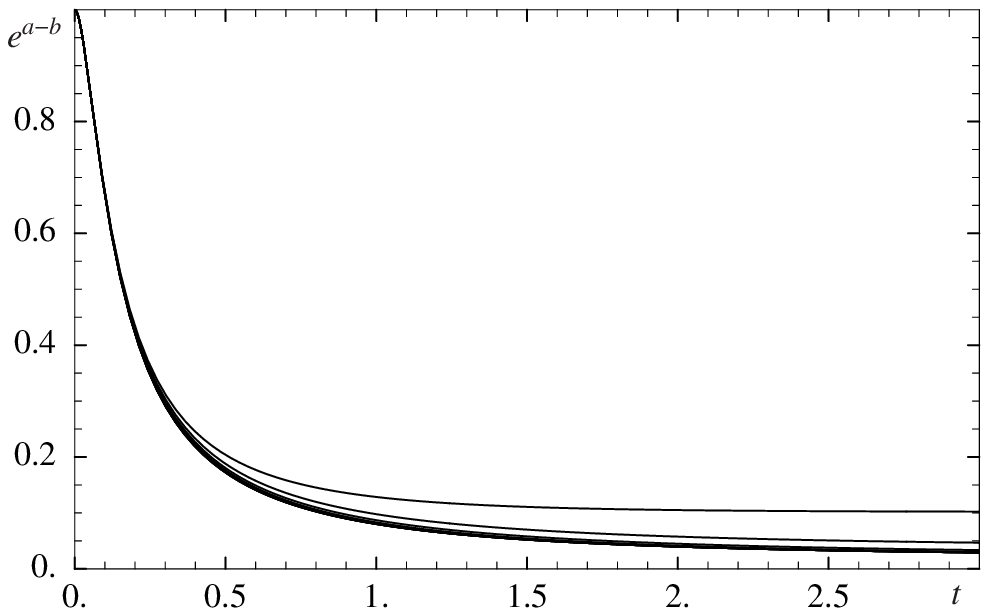}
\caption{Pseudo-eccentricity $e^{a-b}$ for the case $\text{W}\Lambda
  w$ with $\lambda=1$, $\rho_\text{i}=10$,
  $\epsilon_\text{i}=200$. Curves are for $w$ from $-1$ to $1$ with
  step $0.2$ from top to bottom. \label{fig-w-eccentricity-1}}
\end{figure}

\begin{figure}
\includegraphics[width=10cm]{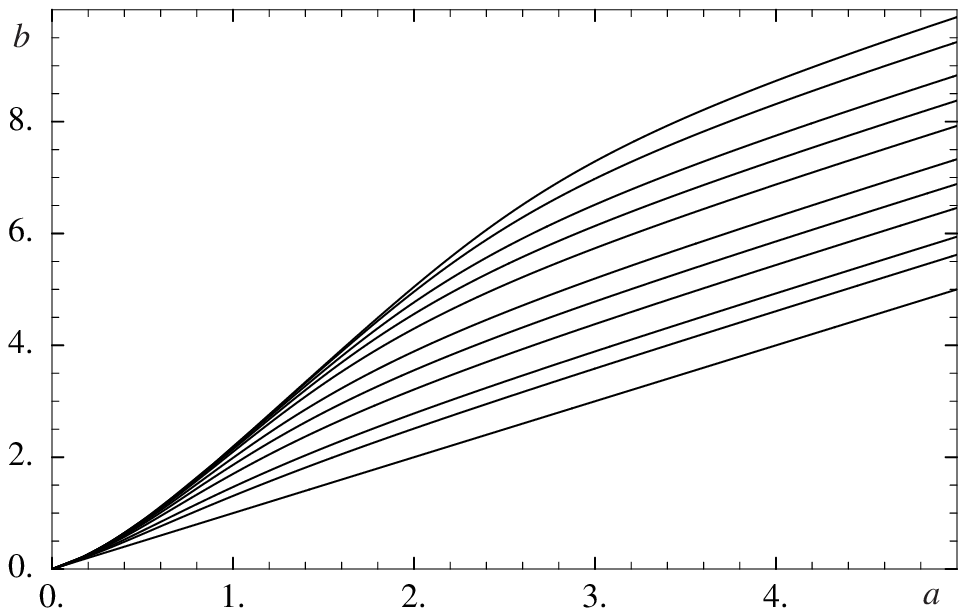}
\caption{Expansion parameters $a$ and $b$ for the case
  $\text{W}\Lambda$ with $\lambda=1$, $\rho_\text{i}=0$. Curves are
  for $\epsilon_\text{i}=0,1,2,5,10,20,50,100,200,500,1000$ from
  bottom to top. \label{fig-w-ab}}
\end{figure}

\begin{figure}
  \includegraphics[width=10cm]{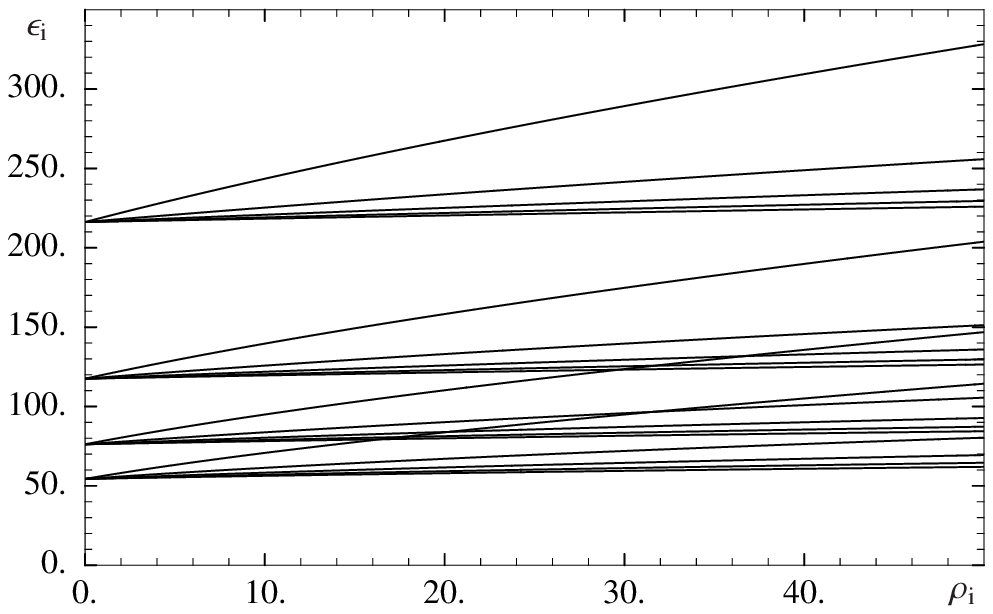}
  \caption{Asymptotic value of the pseudo-eccentricity for the case
    $\text{W}\Lambda w$ with $\lambda=1$ as a function of
    $\rho_\text{i}$ and $\epsilon_\text{i}$. Sets of curves are for
    $\e^{a-b}$ equal to $0.02,0.03,0.04,0.05$ from top to bottom; the
    abscissa corresponds to $\e^{a-b}=1$. Curves in each set are for
    $w$ equal to $-0.5, -0.25, 0, 0.25, 0.5$ from top to bottom.
    \label{fig-w-eccentricity-2}}
\end{figure}

\begin{figure}
  \includegraphics[width=10cm]{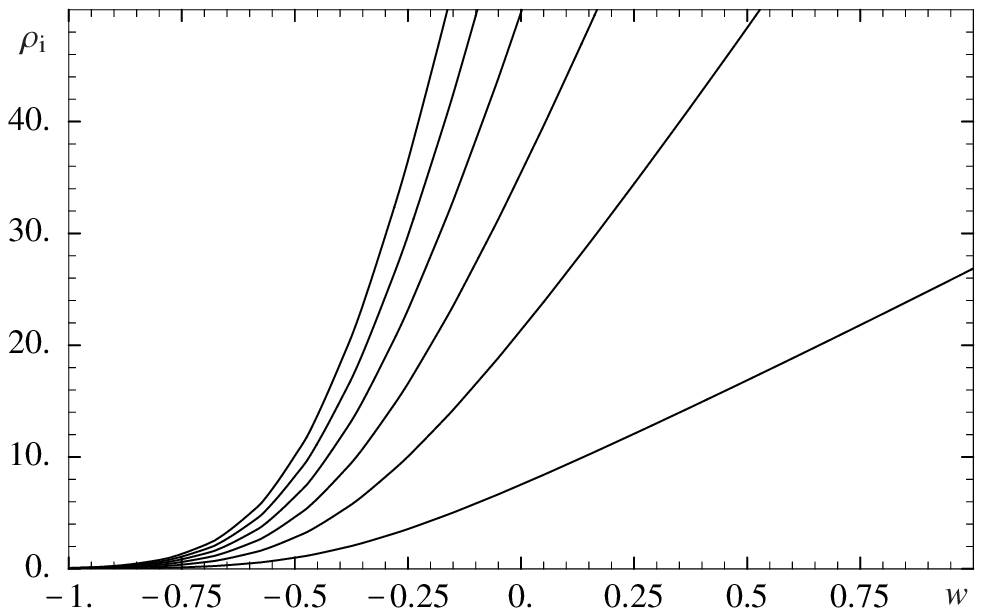}
  \caption{Asymptotic value of the pseudo-eccentricity for the case
    $\text{W}\Lambda w$ with $\lambda=1$ as a function of $w$ and
    $\rho_\text{i}$ for $\epsilon_\text{i}=200$. Curves are for
    $\e^{a-b}$ from $0.0115$ to $0.0120$ with step $0.0001$ from
    bottom to top. \label{fig-w-eccentricity-3}}
\end{figure}


The plots for walls shows a number of qualitative differences with the
previous cases of magnetic fields and strings. The expansion
parameters $a$ and $b$ change with time (see Figs.~17 and 18) more
like the string case, where both grow monotonically; but note that now
$b$ grows faster than $a$, which is the reverse of the behavior seen
for strings. This means the expansion for walls is always prolate,
unlike the expansions for strings and magnetic fields which are always
oblate.

For walls, the matter density, Fig.~19, falls faster with time than
its string and magnetic field counterparts due to the fact that the
overall expansion, and therefore density dilution, is faster for
walls. On the other hand, due to the nature of its contribution to the
stress energy tensor, the wall energy density, Fig.~20, falls more
slowly than the magnetic field and string energy densities.

Figure~21 for the pseudo-eccentricity provides another way of
visualizing the prolateness of the wall expansion, since $e^{a-b}$ is
always less than one in this case. Figure~22 for $a$ versus $b$
represents the degree of prolateness for wall expansion for a variety
of initial conditions. Figure~23 shows asymptotic pseudo-eccentricity
contours for walls as a function of initial matter and wall energy
densities, and Fig.~24 gives asymptotic pseudo-eccentricity contours
for walls as a function of matter density and equation of state
parameter $w$.

\section{Conclusions}\label{S:C}

Einstein's equations for magnetic fields that extend across the
Universe have been considered elsewhere. Examples include
cylindrically symmetric magnetic geons, exact solutions for planar
geometry with magnetic fields and dust, and asymptotics
\cite{magnetic-fields}. None of these studies contain exact solutions
with cosmological constant and magnetic fields ($\Lambda
\text{M}$). We have not only given exact solutions to the $\Lambda
\text{M}$ case, but also have found exact $\Lambda \text{M}$ plus dust
solutions. In addition we have exact solutions when the magnetic
fields are replaced by uniform arrangements of cosmic strings or
cosmic domain walls. Finally we have given approximate solutions in
all these cases where dust can be replaced by matter with an arbitrary
value of $w$ in its equation of state. All our solutions have planar
symmetry.

The cosmic microwave background (CMB) and other modern cosmological
data are of such high quality that it is now possible to study aspects
of the Universe that were previously completely out of reach. In order
to carry out these investigations, it may be necessary to go beyond
the homogeneous isotropic big bang/inflationary cosmology and compare
the data with less symmetric but perhaps more realistic models. In the
case of planar symmetry studied here, an understanding of the density
perturbations and structure formation requires perturbing around
planar symmetric solutions. Here we have taken a step in that
direction by considering a planar symmetric universe with eccentric
expansion, and have shown exact solutions can be obtained even when
the eccentricity is large. This will allow a density perturbation
analysis to be carried out in these cosmologies, which in turn can be
compared with CMB data and galaxy structure and correlation data
\cite{EccIII}.

It is not just a mathematical exercise to consider planar symmetry. We
know that magnetic fields and cosmic defects can be produced in the
early universe. In the case of magnetic fields, their energy density
$\epsilon$ at its production epoch can be a substantial fraction of
the matter density $\rho$, and this can cause spherical symmetry to be
lost in a cosmology. If the typical magnetic domain size $D$ is small
compared to $H^{-1}$, then the local expansion is eccentric, while the
average global expansion remains spherical, while if $D \geq H^{-1}$,
then the whole universe expands eccentrically until $D$ comes within
the horizon. Also, if $D \leq H^{-1}$ initially, a period of inflation
can push regions of size $D$ outside the horizon, and we are again in
a situation of eccentric expansion.

The planar symmetric cases of cosmic strings and domain walls are
somewhat more artificial, since they are assumed to be static and
aligned. However, this may not be totally unrealistic when considered
from the perspective of more fundamental theories. For instance,
certain AdS/CFT theories derived from string theory have parallel
walls, and other theories with branes can have strings connecting
them. If two parallel walls, both outside the horizon, were connected
by strings, then the strings would be expected to be parallel on
average even if they had some dynamics. Based on the above remarks,
and with the knowledge of the fact that aligned walls and strings both
produce planar symmetry, we have given exact solutions for these cases
as well.

Even though magnetic fields, string, and wall systems all have planar
symmetry, the form of their energy momentum tensors differ. For
magnetic fields, ${T^\mu}_\nu$ is traceless, and so this case has
similarities with a radiation filled universe. For strings and walls,
the trace of ${T^\mu}_\nu$ does not vanish, so there are some
similarities with the non-relativistic matter component. Strings and
walls are under tension, so they also have some similarities with
vacuum energy. To see all these properties, we have solved the
equations of motion exactly for many cases of interest. The large-time
behaviors of these solutions are summarized in
Table~\ref{table:summary}.

\begin{table}
\caption{\label{table:summary} Summary of large-time behavior for
various quantities for ten different cases of universe content. For
each choice of an anisotropic component, magnetic fields (M), strings
(S) or walls (W), matter with $w=0$ or with $0<w<1$ is included and
cosmological constant is either present ($\Lambda$) or absent. Only
the leading terms in asymptotics are given and $\tilde
t=(\l/3)^{\frac{1}{2}}t$.}
\begin{ruledtabular}
\begin{tabular}{cccccc}
& $\eps$ & $\rho$ & $\e^a$ & $\e^b$ & $\e^{a-b}$\\ \hline
{\rule[-1ex]{0ex}{5ex} $\text{M}\Lambda w$} & $\e^{-4\tilde t}$ &
$\e^{-3(1+w)\tilde t}$ & $\e^{\tilde t}$ & $\e^{\tilde t}$ & $\ge 1$
\\
{\rule[-1ex]{0ex}{2ex} $\text{M}\Lambda 0$} & $\e^{-4\tilde t}$ &
$\e^{-3\tilde t}$ & $\e^{\tilde t}$ & $\e^{\tilde t}$ & $\ge 1$ \\
{\rule[-1ex]{0ex}{2ex} $\text{M}w$} & $t^{-\frac{8}{3}}$ & $t^{-2}$ &
$t^{\frac{2}{3}}$ & $t^{\frac{2(1-2w)}{3(1+w)}}$ &
$t^{\frac{2w}{1+w}}$ \\
{\rule[-2ex]{0ex}{2ex} $\text{M}0$} & $t^{-\frac{8}{3}}$ & $t^{-2}$
& $t^{\frac{2}{3}}$ & $t^{\frac{2}{3}}$ & $\ge 1$ \\
%
{\rule[-1ex]{0ex}{5ex} $\text{S}\Lambda w$} & $\e^{-2\tilde t}$ &
$\e^{-3(1+w)\tilde t}$ & $\e^{\tilde t}$ & $\e^{\tilde t}$ & $\ge 1$
\\
{\rule[-1ex]{0ex}{2ex} $\text{S}\Lambda 0$} & $\e^{-2\tilde t}$ &
$\e^{-3\tilde t}$ & $\e^{\tilde t}$ & $\e^{\tilde t}$ & $\ge 1$\\
{\rule[-1ex]{0ex}{2ex} $\text{S}w$} & $t^{-2}$ & $t^{-2}$ & $t$
& $t^{-\frac{2w}{1+w}}$ & $t^{\frac{1+3w}{1+w}}$ \\
{\rule[-2ex]{0ex}{2ex} $\text{S}0$} & $t^{-2}$ & $t^{-2}$ & $t$ &
$t^{-2}$ & $\ge 1$ \\ 
%
{\rule[-1ex]{0ex}{5ex} $\text{W}\Lambda 0$} & $\e^{-\tilde t}$ &
$\e^{-3\tilde t}$ & $\e^{\tilde t}$ & $\e^{\tilde t}$ & $\le 1$\\
{\rule[-1ex]{0ex}{2ex}$\text{W}0$} & $t^{-2}$ & $t^{-\frac{10}{3}}$ &
$t^{\frac{2}{3}}$ & $t^2$ & $t^{-\frac{4}{3}}$
\end{tabular}
\end{ruledtabular}
\end{table}

In all these cases, the universe undergoes eccentric expansion and in
some instances eccentric inflation. Our analysis is completely
general, and in order to apply these results, more input is necessary,
e.~g., initial conditions need to be specified, perhaps as derived
from a model with early universe phase transitions. Time scales need
to be fixed, e.~g., when did the phase transition take place? For
instance, for magnetic field production, a phase transition not far
above the electroweak scale may be effective in producing eccentric
effects and at the same time remaining compatible with other
requirements on the cosmological model, e.~g., successful
baryogeneses. If the magnetic field production scale were too high,
then there is a danger that all the eccentric effects could be washed
out.

As stated above, with exact planar symmetric solutions at hand, we are
now in a position to begin density perturbations
analysis~\cite{EccIII}. To apply the results of this paper, it will be
necessary to consider how the spectrum of density perturbations are
effected by asymmetric expansion.  Since perturbations get laid down
by quantum fluctuations and then asymmetrically expanded in our
models, any initial spherical perturbation becomes ellipsoidal. After
a while, the expansion becomes spherically symmetric again, but as
long as perturbations remain outside the horizon they stay
ellipsoidal.  Only after they reenter our horizon will they be able to
adjust (they will probably start to oscillate between prolate and
oblate with frequency that depends on size and overdensity). So if the
perturbations are just entering at last scattering they should be
ellipsoidal. The smaller they are at last scattering, the more they
have oscillated and if damped, the closer to spherical they should
be. Hence, the larger scale perturbations (corresponding to smaller
$\ell$) will have a better memory of the eccentric phase. This would
appear to agree with what seems to be hinted at in the WMAP
observations: more distortion of the low $\ell$ modes. However, a
detailed phenomenological analysis needs to be carried out to confirm
these facts.

To summarize, what we need are modes that expanded eccentrically to be
entering the horizon at the time of last scattering and then to feed
this information into a Sachs-Wolfe type of calculation. This is a
most interesting and challenging calculation, since it requires a full
reanalysis of the density perturbations in eccentric geometry. In this
paper we have moved toward this goal. We have carried out exact
calculations of the evolution of a variety of Universes with
asymmetric matter content. In some cases, namely when $w$ is neither
zero nor minus one, we have been forced to use approximate methods. We
have explored the asymptotic behavior of both the exact and
approximate cases. Our results provide a starting point for the
analysis of density perturbations in asymmetric cosmologies. WMAP and
its successors will be able to either bound or detect effects of
asymmetric inflation and we have taken the first steps in the
theoretical exploration in that direction.

Finally, we make a few comments about the case where there are
multiple magnetic domains within the cosmological horizon. (A similar
discussion would apply to strings and walls.) If the domains are
randomly oriented then what one should expect is eccentric expansion
within each domain, with dependence on the local value of the
cosmological constant, magnetic field strength, and matter
content. Locally there is planar symmetry, but globally the Universe
would look isotropic if averaged over many domains. One effect of the
averaging would be an alteration of the power spectrum on scales of
order of the domain size. This assumes the domains have a preferred
size, that is probably on order of the horizon size when they were
produced, if the associated phase transition was second order, or on
the size of the correlation length at production, if the associated
phase transition was first order. This is in contrast to the density
perturbations produced in inflation that typically have a flat power
spectrum. One would also expect to see polarization effects to survive
in the CMB in an isotropic average of magnetic domains. A detailed
analysis of these effects would take dedicated numerical studies.

\begin{acknowledgments}

This work was supported by US DoE grants DE-FG05-85ER40226 (RVB and
TWK) and DE-FG06-85ER40224 (RVB), and by UK's PPARC (AB).

\end{acknowledgments}

\begin{appendix}

\section{Approximate solutions for $\bm{\Lambda}+\text{M}+\text{MATTER}$
  with arbitrary \lowercase{\bm{$w$}}}\label{S:B0}

We develop a simple approximation by expanding around
$\eps=0$. Eliminating $\rho$ in Eqs.~(\ref{B:f-eq}) and
(\ref{B:rho-eq}), we find
\begin{eqnarray}
2wf\frac{\eps^2 f''-\frac{15}{4}\eps f'+\frac{11}{2}f+2\eps^3}{\eps
f'-\fr{11}{4}f+2\eps^2(\l+\eps)}=(1+w)\left[\eps
f'-\fr{1}{4}(11-3w)f+2(1+w)\eps^2(\l+\eps)\right].
\end{eqnarray}
A power series solution in $\epsilon$ to this equation is
\begin{eqnarray}
f\approx\fr{8}{3}\l\eps^2+\fr{8}{3}\eps_\i^{-\frac{3}{4}}
\left[\rho_\i(\eps/\eps_\i)^{\frac{3}{4}w}+4\eps_\i\right]
\eps^{\frac{11}{4}}-8\eps^3+{\cal
O}\left[\eps^{\frac{1}{4}(15+3w)}\right].\label{B0:f}
\end{eqnarray}
The approximate solution for $\rho(\eps)$ can then be found from
Eqs.~(\ref{B:rho}) and (\ref{B:psi}). [A much faster way to calculate
$\rho$ is to use Eq.~(\ref{B:f-eq}) directly. The result,
$\rho\approx\rho_\i(\eps/\eps_\i)^{\frac{3}{4}(1+w)}$, is unacceptably
inaccurate as is clear from both the exact solution (\ref{B:rho0}) for
$w=0$ and the asymptotic form (\ref{B0:rho-e-asympt}) below for
arbitrary $w$.]  The functions $a(\eps)$ and $b(\eps)$ are given by
Eqs.~(\ref{B:a0}) and (\ref{A5s}). Finally, time dependence of the
above functions $\rho(\eps)$, $a(\eps)$ and $b(\eps)$ can be deduced
from the function $\eps(t)$, which is given implicitly by
\begin{eqnarray}
t-t_\i\approx\fr{1}{4} \int^{\eps_\i}_{\eps}\d\eps\left\{
\fr{1}{3}\l\eps^2+\fr{1}{3}\eps_\i^{-\frac{3}{4}}
\left[\rho_\i(\eps/\eps_\i)^{\frac{3}{4}w}+4\eps_\i\right]
\eps^{\frac{11}{4}}-\eps^3\right\}^{-\frac{1}{2}}\label{B0:integral}
\end{eqnarray}
as it follows from $f=\fr{1}{2}\dot{\eps}^2$ and
Eq.~(\ref{B0:f}). 

Comparing Eqs.~(\ref{B:f0}) and (\ref{B0:f}), we notice that the above
approximate solution becomes exact for $w=0$. In addition, being an
expansion in small $\eps$, the approximate solution gives correct
asymptotics for large $t$. To find the behavior of various quantities
for large $t$, we need the corresponding asymptotic of the integral in
Eq.~(\ref{B0:integral}). When $\l>0$, the integral diverges for small
$\eps$, and so we extract this divergent part first; this results in
\begin{eqnarray}
t-t_\i\approx\fr{1}{4}(3/\l)^{\frac{1}{2}}\ln{(\eps_\i/\eps)}-\tau,
\label{t-asympt}
\end{eqnarray}
where 
\begin{eqnarray}
\tau=\fr{1}{4}\int_{0}^{\eps_\i}
\d\eps\left\{\left(\fr{1}{3}\l\eps^2\right)^{-\frac{1}{2}}
-\left[\fr{1}{3}\l\eps^2+\fr{1}{3}\eps_\i^{-\frac{3}{4}}
(\rho_\i(\eps/\eps_\i)^{\frac{3}{4}w}+4\eps_\i)\eps^{\frac{11}{4}}
-\eps^3\right] ^{-\frac{1}{2}}\right\}.\label{B0:tau}
\end{eqnarray}
Similarly extracting the divergent part of $\psi$ for small $\eps$, we
find
\begin{eqnarray}
\rho\sim(\eps/\eps_\i)^{\frac{3}{8}(1+w)}\e^{(1+w)\phi}
\left[1+F(0)\right]^{-1},
\label{B0:rho-e-asympt}
\end{eqnarray}
where
\begin{eqnarray}
&&\phi=-\fr{3}{8}(\eps_\i/\l)+\int_{0}^{\eps_\i}
\d\eps\,\eps(\l+\eps)\left\{\left(\fr{8}{3}\l\eps^2\right)^{-1}
-f^{-1}\right\},\label{B0:phi}\\
&&F(0)=(1+w)\rho_\i\eps_\i\int_{0}^{\eps_\i}
\d\eps\,(\eps/\eps_\i)^{1+\frac{3}{8}(1+w)}\psi f^{-1}.\label{B0:F}
\end{eqnarray}
 
Finally, this results in the following asymptotics:
\begin{eqnarray}
&&\eps\sim \eps_\i\,\e^{-4\sigma},
\label{B0:e-asympt}\\ &&a\sim\sigma,\label{B0:a-asympt}\\ &&b\sim
\sigma-\phi+(1+w)^{-1}\ln\left[1+F(0)\right],
\label{B0:b-asympt}\\
&&\rho\sim\rho_\i [1+F(0)]^{-1} \exp\left[-(1+w)(3\sigma-\phi)\right],
\label{B0:rho-asympt}
\end{eqnarray}
where $\sigma(t)=(\l/3)^{\frac{1}{2}}(t-t_\i+\tau)$.  For large $t$,
both scale factors grow linearly (as in the isotropic case driven by
the cosmological constant only). Due to anisotropy introduced by the
magnetic fields, however, the space has expanded more transversally
than longitudinally. This difference is characterized by the
pseudo-eccentricity whose asymptotic form in this case is
\begin{eqnarray}
\e^{a-b}\sim\e^{\phi}\left[1+F(0)\right]^{-1/(1+w)}.\label{B0:ecc-asympt}
\end{eqnarray}

When $\l=0$, the asymptotics depend on the range of the parameter $w$;
in the most interesting case, $0<w<1$, they are:
\begin{eqnarray}
&&\eps\sim 3^{-\frac{4}{3}}\eps_\i^{-\frac{1}{3}}t^{-\frac{8}{3}},\\
&&a\sim\fr{2}{3}\ln{(\eps_\i^\frac{1}{2}t)},\\ &&b\sim
\frac{2(1-2w)}{3(1+w)}\ln{(\eps_\i^\frac{1}{2}t)},\\ &&\rho\sim
\frac{4(1-w)}{3(1+w)}t^{-2}.
\end{eqnarray}
Thus, in the absence of constant negative pressure from the
cosmological constant, anisotropy causes the space to infinitely
expand in the transverse directions and infinitely contract in the
longitudinal direction. This results in pseudo-eccentricity diverging
for large $t$: $\e^{a-b}\sim (\eps_\i^\frac{1}{2}t)^{2w/(1+w)}$.  In
the case of dust ($w=0$), the asymptotic value for the
pseudo-eccentricity is finite, in agreement with
Eq.~(\ref{B:ecc0-asympt}).

\section{Approximate solutions for $\bm{\Lambda}+\text{S}+\text{MATTER}$
  with arbitrary \lowercase{\bm{$w$}}}\label{S:S0}

Eliminating $\rho$ in Eqs.~(\ref{S:f-eq}) and (\ref{S:rho-eq}), we
find
\begin{eqnarray}
2wf\frac{\eps^2 f''-\fr{9}{2}\eps f'+7f+\eps^3}{\eps
f'-\fr{7}{2}f+\eps^2(\l+\eps)}=(1+w)\left[\eps
f'-\fr{1}{2}(7-3w)f+(1+w)\eps^2(\l+\eps)\right].
\end{eqnarray}
A power series solution to this equation is
\begin{eqnarray}
f=\fr{2}{3}\l\eps^2+2\eps^3+\fr{2}{3}\eps_\i^{-\frac{3}{2}}
\left[\rho_\i(\eps/\eps_\i)^{\frac{3}{2}w}-2\eps_\i\right]
\eps^{\frac{7}{2}}+{\cal
O}\left[\eps^{\frac{1}{2}(9+3w)}\right]\label{S0:f}
\end{eqnarray}
The approximate solution for $\rho$ can then be found from
Eqs.~(\ref{S:rho}) and (\ref{S:psi}). [As in the previous section, a
simple expression
$\rho\approx\rho_\i(\eps/\eps_\i)^{\frac{3}{2}(1+w)}$, which follows
directly from Eq.~(\ref{B:f-eq}), is a poor approximation.] Similar to
the case of magnetic fields, the approximate solution~(\ref{S0:f})
becomes exact for $w=0$.

Proceeding similarly to Appendix~\ref{S:B0}, we find the following
large-time asymptotics:
\begin{eqnarray}
&&\eps\sim
\eps_\i\,\e^{-2\sigma},
\label{S0:e-asympt}\\ &&a\sim\sigma ,\label{S0:a-asympt}\\ &&b\sim
\sigma-\phi +(1+w)^{-1}\ln\left[1+F(0)\right],
\label{S0:b-asympt}\\
&&\rho\sim\rho_\i [1+F(0)]^{-1} \exp\left[-(1+w)(3\sigma-\phi)\right],
\label{S0:rho-asympt}\\
&&\e^{a-b}\sim\e^{\phi}\left[1+F(0)\right]^{-1/(1+w)},\label{S0:ecc-asympt}
\end{eqnarray}
where $\sigma(t)=(\l/3)^{\frac{1}{2}}(t-t_\i+\tau)$ and
\begin{eqnarray}
&&\tau=\fr{1}{2}\int_{0}^{\eps_\i}
\d\eps\left\{\left(\fr{1}{3}\l\eps^2\right)^{-\frac{1}{2}}
-\left[\fr{1}{3}\l\eps^2+\eps^3+\fr{1}{3}\eps_\i^{-\frac{3}{2}}
(\rho_\i(\eps/\eps_\i)^{\frac{3}{2}w}-2\eps_\i)\eps^{\frac{7}{2}}\right]
^{-\frac{1}{2}}\right\},\label{S0:tau}\\
&&\phi=-\fr{3}{4}(\eps_\i/\l)+\fr{1}{2}\int_{0}^{\eps_\i}
\d\eps\,\eps(\l+\eps)\left\{\left(\fr{2}{3}\l\eps^2\right)^{-1}
-f^{-1}\right\},\label{S0:phi}\\
&&F(0)=\fr{1}{2}(1+w)\rho_\i\eps_\i\int_{0}^{\eps_\i}
\d\eps\,(\eps/\eps_\i)^{1+\frac{3}{4}(1+w)}\psi f^{-1}.\label{S0:F}
\end{eqnarray}

In the case of zero vacuum energy, the asymptotics depend on the range
of the parameter $w$; in the most interesting case, $0<w<1$, they are:
\begin{eqnarray}
&&\eps\sim t^{-2},\\ &&a\sim\ln{(\eps_\i^\frac{1}{2}t)},\\ &&b\sim
-\frac{2w}{1+w}\ln{(\eps_\i^\frac{1}{2}t)},\\ &&\rho\sim
\frac{2(1-w)}{1+w}t^{-2}.
\end{eqnarray}
As in the magnetic field case, the pseudo-eccentricity diverges for
large $t$: $\e^{a-b}\sim (\eps_\i^\frac{1}{2}t)^{(1+3w)/(1+w)}$.

\end{appendix}


\begin{thebibliography}{99}



\bibitem{Guth:1980zm}
  A.~H.~Guth,
  Phys.\ Rev.\ D {\bf 23}, 347 (1981).

\bibitem{Linde:1981mu}
  A.~D.~Linde,
  Phys.\ Lett.\ B {\bf 108}, 389 (1982).

\bibitem{Albrecht:1982wi}
  A.~Albrecht and P.~J.~Steinhardt,
  Phys.\ Rev.\ Lett.\  {\bf 48}, 1220 (1982).

\bibitem{Linde:1983gd}
  A.~D.~Linde,
  Phys.\ Lett.\ B {\bf 129}, 177 (1983).

\bibitem{Liddle:1993fq}
  A.~R.~Liddle and D.~H.~Lyth,
  Phys.\ Rept.\  {\bf 231}, 1 (1993)
  [arXiv:astro-ph/9303019].

\bibitem{Guth:1982ec}
  A.~H.~Guth and S.~Y.~Pi,
  Phys.\ Rev.\ Lett.\  {\bf 49}, 1110 (1982).

\bibitem{Kodama:1985bj}
  H.~Kodama and M.~Sasaki,
  Prog.\ Theor.\ Phys.\ Suppl.\  {\bf 78} (1984) 1.

\bibitem{Mukhanov:1990me}
  V.~F.~Mukhanov, H.~A.~Feldman and R.~H.~Brandenberger,
  Phys.\ Rept.\  {\bf 215}, 203 (1992).

\bibitem{Peebles} P.~J.~E.~Peebles, \emph{The large-scale structure of
the universe,} Princeton University Press, Princeton, 1980.

\bibitem{Padmanabhan} T.~Padmanabhan, \emph{Structure formation in the
universe}, Cambridge University Press, Cambridge, 1993.

\bibitem{Smoot:1992td}
  G.~F.~Smoot {\it et al.},
  Astrophys.\ J.\  {\bf 396}, L1 (1992).

\bibitem{Bennett:1996ce}
  C.~L.~Bennett {\it et al.},
  Astrophys.\ J.\  {\bf 464}, L1 (1996)
  [arXiv:astro-ph/9601067].

\bibitem{Kogut:1996us}
A.~Kogut {\it et. al},
arXiv:astro-ph/9601060.

\bibitem{Lyth:1998xn}
  D.~H.~Lyth and A.~Riotto,
  Phys.\ Rept.\  {\bf 314}, 1 (1999)
  [arXiv:hep-ph/9807278].

\bibitem{Linde} A.~D.~Linde, {\it Particle physics and inflationary
cosmology}, Harwood Academic Publishers, Chur, Switzerland, 1990
[arXiv:hep-th/0503203]

\bibitem{KT} E. W. Kolb and M. S. Turner, {\it The early universe},
Addison-Wesley, Reading, MA, 1990.

\bibitem{Dodelson} S. Dodelson, \emph{Modern cosmology,} Academic
Press, San Diego, 2003.

\bibitem{KR} M.~Yu.~Khlopov and S.~G.~Rubin, \emph{Cosmological
Pattern of Microphysics in the Inflationary Universe}, Springer, 2004.

\bibitem{Berera:2003tf} A.~Berera, R.~V.~Buniy and T.~W.~Kephart,
JCAP {\bf 10}, 016 (2004).

\bibitem{Bennett:2003bz} C.~L.~Bennett {\it et al.},
Astrophys.\ J.\ Suppl.\ {\bf 148}, 1 (2003).

\bibitem{Spergel:2003cb}
D.~N.~Spergel {\it et al.},
Astrophys.\ J.\ Suppl.\  {\bf 148}, 175 (2003).

\bibitem{Hinshaw:2003ex}
  G.~Hinshaw {\it et al.},
  Astrophys.\ J.\ Suppl.\  {\bf 148}, 135 (2003)
  [arXiv:astro-ph/0302217].

\bibitem{Tegmark:2003ve}
  M.~Tegmark, A.~de Oliveira-Costa and A.~Hamilton,
  Phys.\ Rev.\ D {\bf 68}, 123523 (2003)
  [arXiv:astro-ph/0302496].

\bibitem{deOliveira-Costa:2003pu}
A.~de Oliveira-Costa, M.~Tegmark, M.~Zaldarriaga and A.~Hamilton,
arXiv:astro-ph/0307282.

\bibitem{Berera:1997wz}
  A.~Berera, L.~Z.~Fang and G.~Hinshaw,
  Phys.\ Rev.\ D {\bf 57}, 2207 (1998)
  [arXiv:astro-ph/9703020].

\bibitem{Berera:2000wz}
  A.~Berera and A.~F.~Heavens,
  Phys.\ Rev.\ D {\bf 62}, 123513 (2000)
  [arXiv:astro-ph/0010366].

\bibitem{Kronberg:1993vk}
  P.~P.~Kronberg,
  Rept.\ Prog.\ Phys.\  {\bf 57}, 325 (1994).

\bibitem{Hindmarsh:1994re}
  M.~B.~Hindmarsh and T.~W.~B.~Kibble,
  Rept.\ Prog.\ Phys.\  {\bf 58}, 477 (1995)
  [arXiv:hep-ph/9411342].

\bibitem{Wick:2000yc}
  S.~D.~Wick, T.~W.~Kephart, T.~J.~Weiler and P.~L.~Biermann,
  Astropart.\ Phys.\  {\bf 18}, 663 (2003)
  [arXiv:astro-ph/0001233].

\bibitem{Taub:1950ez} A.~H.~Taub,
Annals Math.\  {\bf 53}, 472 (1951).

\bibitem{magnetic-fields} M.~A.~Melvin, Phys.\ Lett.\ {\bf 8}, 65
  (1964); Ya.~B.~Zeldovich and I.~D.~Novikov, \emph{Relativistic
  astrophysics,} vol. 2, University of Chicago Press, Chicago (1983);
  J.~D.~Barrow and R.~Maartens, Phys.\ Rev.\ D {\bf 59}, 043502 (1999)
  [arXiv:astro-ph/9808268].

\bibitem{Savvidy:1977as} G.~K.~Savvidy, 
  Vacuum State Of Gauge Theories And Asymptotic 
  Lett.\ B {\bf 71}, 133 (1977).  

\bibitem{Vachaspati:1991nm}
  T.~Vachaspati,
  Phys.\ Lett.\ B {\bf 265}, 258 (1991).

\bibitem{Enqvist:1994rm}
  K.~Enqvist and P.~Olesen,
  Phys.\ Lett.\ B {\bf 329}, 195 (1994)
  [arXiv:hep-ph/9402295].

\bibitem{Berera:1998hv}
  A.~Berera, T.~W.~Kephart and S.~D.~Wick,
  Phys.\ Rev.\ D {\bf 59}, 043510 (1999)
  [arXiv:hep-ph/9809404].

\bibitem{Birch:1982}
P.~Birch, Nature, {\bf 298}, 451 (1982)

\bibitem{Nodland:1997cc}
  B.~Nodland and J.~P.~Ralston,
  Phys.\ Rev.\ Lett.\  {\bf 78}, 3043 (1997)
  [arXiv:astro-ph/9704196].

\bibitem{Ralston:2003pf}
  J.~P.~Ralston and P.~Jain,
  Int.\ J.\ Mod.\ Phys.\ D {\bf 13}, 1857 (2004)
  [arXiv:astro-ph/0311430].

\bibitem{Jaffe:2005pw}
  T.~R.~Jaffe, A.~J.~Banday, H.~K.~Eriksen, K.~M.~Gorski and F.~K.~Hansen,
  arXiv:astro-ph/0503213.

\bibitem{Hutsemekers:2005iz}
  D.~Hutsemekers, R.~Cabanac, H.~Lamy and D.~Sluse,
  arXiv:astro-ph/0507274.

\bibitem{Cornish:1997ab}
  N.~J.~Cornish, D.~N.~Spergel and G.~D.~Starkman,
  Class.\ Quant.\ Grav.\  {\bf 15}, 2657 (1998)
  [arXiv:astro-ph/9801212].

\bibitem{Luminet:2003dx}
  J.~P.~Luminet, J.~Weeks, A.~Riazuelo, R.~Lehoucq and J.~P.~Uzan,
  Nature {\bf 425}, 593 (2003)
  [arXiv:astro-ph/0310253].

\bibitem{Cornish:2003db}
  N.~J.~Cornish, D.~N.~Spergel, G.~D.~Starkman and E.~Komatsu,
  Phys.\ Rev.\ Lett.\  {\bf 92}, 201302 (2004)
  [arXiv:astro-ph/0310233].

\bibitem{Schwarz:2004gk}
  D.~J.~Schwarz, G.~D.~Starkman, D.~Huterer and C.~J.~Copi,
  Phys.\ Rev.\ Lett.\  {\bf 93}, 221301 (2004)
  [arXiv:astro-ph/0403353].

\bibitem{Kolb:2005me}
 D.~H.~Coule,
 Class.\ Quant.\ Grav.\  {\bf 22}, R125 (2005)
 [arXiv:gr-qc/0412026];
 E.~W.~Kolb, S.~Matarrese, A.~Notari and A.~Riotto,
 arXiv:hep-th/0503117;
 L.~Knox,
 arXiv:astro-ph/0503405;
 D.~f.~Zeng and Y.~h.~Gao,
 arXiv:hep-th/0503154;
 D.~L.~Wiltshire,
 arXiv:gr-qc/0503099;
 G.~Geshnizjani, D.~J.~H.~Chung and N.~Afshordi,
 arXiv:astro-ph/0503553;
 E.~E.~Flanagan,
 Phys.\ Rev.\ D {\bf 71}, 103521 (2005)
 [arXiv:hep-th/0503202];
 C.~M.~Hirata and U.~Seljak,
 arXiv:astro-ph/0503582;
 A.~Notari,
 arXiv:astro-ph/0503715;
 J.~W.~Moffat,
 arXiv:astro-ph/0504004;
 S.~Rasanen,
 arXiv:astro-ph/0504005.
 B.~M.~N.~Carter, B.~M.~Leith, S.~C.~C.~Ng, A.~B.~Nielsen and
D.~L.~Wiltshire,
 arXiv:astro-ph/0504192;
 S.~P.~Patil,
 arXiv:hep-th/0504145;
 E.~R.~Siegel and J.~N.~Fry,
 arXiv:astro-ph/0504421;
 A.~A.~Coley, N.~Pelavas and R.~M.~Zalaletdinov,
 arXiv:gr-qc/0504115.
 P.~Martineau and R.~H.~Brandenberger,
perturbations,''
 arXiv:astro-ph/0505236;
 J.~W.~Moffat,
 arXiv:astro-ph/0505326;
 M.~Giovannini,
 arXiv:hep-th/0505222;
 D.~f.~Zeng and Y.~h.~Gao,
 arXiv:gr-qc/0506054;
 V.~F.~Cardone, A.~Troisi and S.~Capozziello,
 arXiv:astro-ph/0506371;
 H.~Alnes, M.~Amarzguioui and O.~Gron,
 arXiv:astro-ph/0506449;
 E.~W.~Kolb, S.~Matarrese and A.~Riotto,
 arXiv:astro-ph/0506534;
 D.~f.~Zeng and H.~j.~Zhao,
 arXiv:gr-qc/0506115;
 M.~Giovannini,
 arXiv:astro-ph/0506715;
  M.~Jankiewicz and T.~W.~Kephart,
  arXiv:hep-ph/0510009.

\bibitem{EccIII}
A.~Berera, R.~V.~Buniy and T.~W.~Kephart,
``The eccentric universe: density perturbations,'' in preparation.

\bibitem{energy-momentum} S.~W.~Hawking and G.~F.~R.~Ellis, \emph{The
Large Scale Structure of Space-Time}, Cambridge University Press,
Cambridge, 1973.

\bibitem{nec} The dominant energy condition (DEC) states that
  $T_{\mu\nu}t^\mu t^\nu\ge 0$ and $T_{\mu\nu}t^\mu$ is timelike or
  null for all timelike $t^\mu$. The null energy condition (NEC)
  states that $T_{\mu\nu}n^\mu n^\nu\ge 0$ for all null
  $n^\mu$. Clearly, if the NEC is violated, than the DEC is also
  violated. It can be shown that for a broad class of models,
  violation of the NEC leads to instability; see R.~V.~Buniy and
  S.~D.~H.~Hsu,
arXiv:hep-th/0502203; to appear in Phys.~Lett.~B.  

\bibitem{Weinberg} S.~Weinberg, \emph{Gravitation and Cosmology:
Principles and Applications of the General Theory of Relativity}, John
Wiley and Sons, New York, 1972.

\bibitem{Landau} L.~D.~Landau and E.~M.~Lifshitz, \emph{Quantum
Mechanics: Non-relativistic Theory}, Pergamon Press, Oxford, 1977.



\end{thebibliography}
\end{document}